\documentclass[twocolumn,secnumarabic,amsmath,amssymb,
nobibnotes, aps, prb,groupedaddress,superscriptaddress]{revtex4-2}
\usepackage{setspace}
\usepackage[hidelinks]{hyperref}

\usepackage{graphicx}
\usepackage{dcolumn}

\usepackage{mathtools}
\usepackage{color}
\usepackage{float}
\usepackage{booktabs}
\usepackage[english]{babel}
\usepackage[utf8]{inputenc}
\usepackage{enumerate}
\usepackage[cmyk,dvipsnames]{xcolor}
\usepackage[normalem]{ulem}
\usepackage{multirow,tabularx}

\usepackage{algorithm}
\usepackage{algpseudocode}
\algnewcommand{\LineComment}[1]{\State // #1}

\newcommand{\fccIr}{fcc-Pd/Fe/Ir(111) }

\newcommand{\fccRh}{fcc-Pd/Fe/Rh(111) }

\newcommand{\mmdot}[2]{\mathbf{m}_{#1}\cdot\mathbf{m}_{#2}}
\newcommand{\mmcross}[2]{\mathbf{m}_{#1}\times\mathbf{m}_{#2}}
\newcommand{\mvec}[1]{\mathbf{m}_{#1}}

\newcommand{\iceland}{Science Institute and Faculty of Physical Sciences, University of Iceland, VR-III, 107 Reykjav\'{i}k, Iceland}
\newcommand{\kalmar}{Department of Physics and Electrical Engineering, Linnaeus University, SE-39231 Kalmar, Sweden}
\newcommand{\kiel}{Institut f\"ur Theoretische Physik und Astrophysik,
Christian-Albrechts-Universit\"at zu Kiel, D-24098 Kiel, Germany}
\newcommand{\toulouse}{CEMES, Universit\'e de Toulouse, CNRS, 29 rue Jeanne Marvig, F-31055 Toulouse, France}

\begin{document}

\title{Impact of higher-order exchange on the lifetime of skyrmions and antiskyrmions}
\author{Hendrik Schrautzer}
\thanks{These authors contributed equally to this work.}
\affiliation{\iceland}
\author{Moritz A. Goerzen}
\thanks{These authors contributed equally to this work.}
\affiliation{\toulouse}
\affiliation{\kiel}
\author{Bjarne Beyer}
\affiliation{\kiel}
\author{Soumyajyoti Haldar}
\affiliation{\kiel}
\author{Pavel F. Bessarab}
\affiliation{\iceland}
\affiliation{\kalmar}
\author{Stefan Heinze}
\email[Corresponding author: ]{heinze@physik.uni-kiel.de}
\affiliation{\kiel}
\affiliation{Kiel Nano, Surface, and Interface Science (KiNSIS), University of Kiel, Germany}
\date{\today}

\begin{abstract}
Reliable control of skyrmion lifetime is essential for realizing spintronic devices, yet the role of higher-order exchange -- which can lead to skyrmion stabilization -- remains largely unexplored.
Here we calculate lifetimes of isolated skyrmions and antiskyrmions at transition-metal interfaces based on an atomistic spin model that includes all fourth-order exchange terms.
Within harmonic transition-state theory, we evaluate both energetic and entropic contributions and find substantially enhanced lifetimes when higher-order exchange is included.
The four-spin four-site interaction raises the energy barrier and lowers the curvature of the energy landscape at the collapse saddle point, increasing the pre-exponential factor.
We show that skyrmions and antiskyrmions can remain thermally stable even without Dzyaloshinskii–Moriya interaction (DMI), and that tuning the four-spin term by a small amount modulates the prefactor over orders of magnitude.
Our results identify higher-order exchange as a promising route to stabilize topological magnetic textures -- in particular in systems lacking DMI -- and to engineer their thermally activated decay.
\end{abstract}

\maketitle

\section*{}\label{sec:introduction}
\noindent{\large{\textbf{Introduction}}}\par
\noindent Magnetic skyrmions~\cite{bogdanov1989} and antiskyrmions~\cite{nayak2017} are nanoscale, topologically non-trivial magnetic textures~\cite{nagaosa2013} that have garnered intense interest for their potential use in future spintronic technologies.
Their small size, metastability, and efficient current-driven motion~\cite{iwasaki2013,sampaio2013,zhou2014,woo2016,fert2017} make them attractive candidates for information storage~\cite{fert2013}, logic devices, and neuromorphic computing architectures~\cite{pinna2018, song2020,grollier2020}.
These magnetic textures have been observed in various systems, including non-centrosymmetric bulk magnets~\cite{muehlbauer2009,yu2010}, ultrathin transition-metal films~\cite{heinze2011,romming2013,meyer2019,muckel2021}, multilayers~\cite{moreau2016,boulle2016,soumyanarayanan2017}, and two-dimensional (2D) van der Waals magnets \cite{han2019topological,ding2019observation,wu2020neel,yang2020creation,wu2021van,liu2024magnetic, kartsev2020, Ni2021, xu2022, li2023}. 
A key mechanism enabling their stabilization is the Dzyaloshinskii-Moriya interaction (DMI)~\cite{dzyaloshinskii1957,moriya1960}, which emerges from spin-orbit coupling in systems with broken inversion symmetry such as surfaces~\cite{bode2007}.
In addition to DMI, frustrated Heisenberg exchange interactions -- arising from competing ferromagnetic and antiferromagnetic couplings beyond nearest neighbors -- have been shown to stabilize isolated skyrmions~\cite{leonov2015,lin2016,malottki2017,zhang2017,desplat2019} and antiskyrmions \cite{goerzen2023} even in the absence of DMI.

Recently, higher-order exchange interactions (HOI) have come into focus as another interaction influencing skyrmion stability~\cite{heinze2011,paul2020,xu2022,li2023}.
These exchange interactions, which couple more than two spins, can affect the magnetic ground-state of the system~\cite{kurz2001}.
They can be important for magnetic insulators~\cite{koebler1996,koebler2001} but also for itinerant magnets~\cite{mryasov1996,kurz2001,Lounis2010} and can lead to complex spin structures as observed experimentally~\cite{heinze2011,yoshida2012,Kroenlein2018,romming2018,Spethmann2020,gutzeit2022}.
Extending the pairwise Heisenberg model within perturbation theory~\cite{takahashi1977,macdonald1988,hoffmann2020} gives rise to higher-order energy contributions. 
In this work, we focus on the fourth-order HOI term:
\begin{equation}
\begin{split}
    E_{\text{4-spin}}= -\sum_{ijkl}C_{ijkl}(\mmdot ij) (\mmdot kl)~,
\label{eq:hamiltonian_HO_general}
\end{split}
\end{equation}
including up to four different magnetic moments $\mathbf{m}_i$ (4-spin) of the system.
However, if  $i=k$ and $j=l$, effectively only two lattice sites (2-site) are contributing to the so called biquadratic or 4-spin-2-site interaction. 
Similar 4-spin-3-site and 4-spin-4-site terms have been discussed~\cite{hoffmann2020}.
In this work these are referred to as the biquadratic, the 3-site and the 4-site HOI terms.

A critical challenge in information technologies based on magnetic textures is ensuring the stability of these nanoscale bits against thermal fluctuations and thus maximizing their lifetime $\tau$ given by the Arrhenius law~\cite{malottki2019,varentcova2020, goerzen2024thesis}:
\begin{align}
    \tau^{-1} = \nu_0\textrm{e}^{-\beta\Delta E}~,
    \label{eq:arrhenius}
\end{align}
where $\beta=1/(k_{\text{B}} T)$ with the temperature $T$.
This does not only include maximizing the corresponding energy barrier $\Delta E$.
It has been shown that entropic and dynamic contributions -- incorporated in the pre-exponential factor $\nu_0$ -- can significantly affect the lifetime, leading to differences of several orders of magnitude even for systems with similar energy barriers~\cite{hagemeister2015,wild2017,desplat2018,malottki2019}.

The role of HOI in stabilizing skyrmions and antiskyrmions has only recently been appreciated.
In particular, it has been demonstrated that the 4-site interaction can significantly alter the energy landscape enhancing the energy barrier of topological spin structures~\cite{paul2020}.
However, despite the importance of the entropic and dynamic contributions to lifetimes of magnetic textures, the effect of HOI on the pre-exponential factor -- essential to quantify the stability of metastable spin states via their lifetime -- has not been investigated yet.

In this work, we present the development of a general computational framework for lifetime calculations of topological spin states based on transition-state theory including higher-order exchange interactions and we use it to study lifetimes of  skyrmions and antiskyrmions in ultrathin films.
An efficient implementation of HOI enables us to investigate how it modifies the curvature of the energy landscape and thereby affects entropic contribution to the stability of magnetic textures. 
We apply our approach to study topological spin states in the famous ultrathin film system Pd/Fe/Ir(111), in which isolated skyrmions have been experimentally first observed~\cite{romming2013,hagemeister2015,muckel2021}, and the related system Pd/Fe/Rh(111), in which nanoscale skyrmions have been predicted based on first-principles calculations~\cite{haldar2018}.

Using harmonic transition-state theory and an atomistic spin model parameterized via density functional theory (DFT), we quantify the impact of HOI on the lifetimes of isolated skyrmions and antiskyrmions.
We show that the pre-exponential factor in the Arrhenius law can be modified by orders of magnitude for small changes of the four-site four spin interaction within the
range of typical values. 
Our findings highlight the potential role of HOI in determining the thermal robustness of topological spin textures and provide key insights for designing stable, nanoscale magnetic bits for device applications.
This is especially interesting in light of the advent of defect-free 2D van-der-Waals magnets, which often exhibit inversion symmetry in their structure and therefore lack DMI but may exhibit significant HOI.

\begin{figure*}
    \centering
    \includegraphics[width=1.0\linewidth]{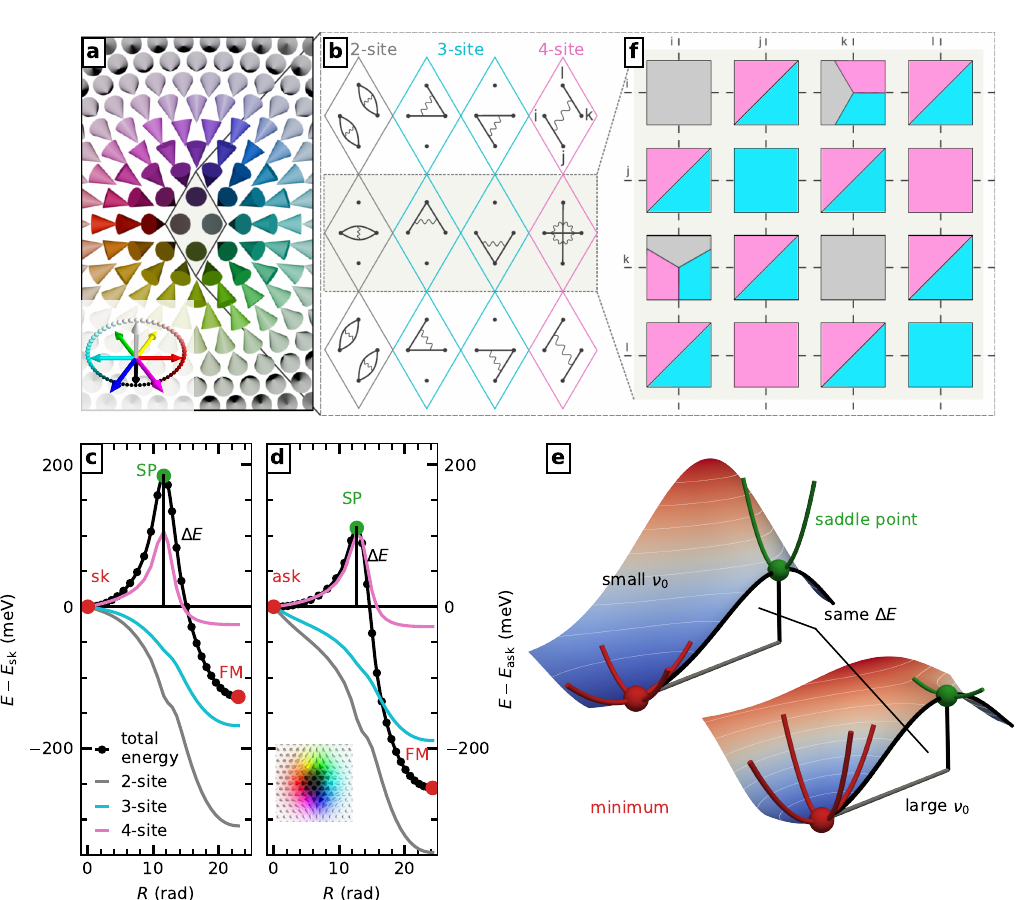}
    \caption{\textbf{Higher-order exchange interactions and harmonic transition state theory}. \textbf{a} Meta-stable skyrmion at $B_\perp-B_C=3.95$~T in \fccIr including higher-order exchange interactions (HOI). The color-code for the orientation of magnetic moments used in this work is shown in the lower-left corner.
    \textbf{b} The interactions of magnetic moments due to the biquadratic (2-site, gray),  the 3-site (cyan) and the 4-site (pink) HOI are schematically depicted by straight lines (products $\mathbf{m}_i\cdot\mathbf{m}_j$) and curved lines (combinations of products) (see Eq.~(\ref{eq:hamiltonian})).  \textbf{c,d} Total energy including all terms in Eq.~(\ref{eq:hamiltonian}) (black) and HOI energy contributions (gray, cyan, pink) to the total energy as a function of the geodesic distance $R$ along the minimum energy path (MEP) of a skyrmion (\textbf{c}) and antiskyrmion annihilation (\textbf{d}) at $B_\perp-B_C=3.95$~T. The minima (saddle points (SPs)) are visualized by red (green) circles and the energy barrier is depicted by a vertical black line. \textbf{e} Schematic two-dimensional energy landscapes with low curvature at the minimum (red) and large curvature at the SP (green) in the upper left corner and vice versa in the lower right corner. \textbf{f} Schematic adjacency pattern for a sub-matrix of the matrix of second-order derivatives for four magnetic moments ($\mathbf{m}_i$, $\mathbf{m}_j$, $\mathbf{m}_k$, $\mathbf{m}_l$) in nearest neighbor distance of each other. For the mid row of panel \textbf{b}
    chosen as an example, each block is colored according to the specific HOI term contributing to its values.
    }
    \label{fig:concept_barrier_vs_curvature}
\end{figure*}

\section*{}\label{sec:results}
\noindent{\large{\textbf{Results}}}\par
\noindent{\textbf{Atomistic spin simulations.}} 
We investigate the stability of skyrmions (Fig.~\ref{fig:concept_barrier_vs_curvature}a) and antiskyrmions (inset of Fig.~\ref{fig:concept_barrier_vs_curvature}d) in \fccIr and \fccRh using an atomistic spin model parameterized via DFT calculations (see Supplementary Note 1).
Note, that fcc indicates the stacking order of the Pd overlayer.
In this spin model, the magnetic moments are represented as classical unit vectors $\mathbf{m}_i$ located on each site $i$ of a 2D hexagonal lattice.
The total energy of a magnetic configuration is given by the Hamiltonian 
\begin{widetext}
\begin{align}
    \label{eq:hamiltonian}
    H =&  -\sum_{i,j}J_{ij} (\mmdot ij) 
    -\sum_{i,j}\mathbf{D}_{ij} \cdot( \mmcross ij)
    - K_{\rm u} \sum_{i} (\mathbf{m}_i\cdot\mathbf{\hat{e}}_z)^2
    -\mu\sum\limits_{i}\mathbf{m}_i\cdot\mathbf{B}_\text{ext} 
    -\sum_{ij} B_{ij} ( \mvec i \cdot \mvec j)^2 \notag\\
    &-2 \sum_{ijk} Y_{ijk} (\mmdot ij )(\mmdot ik)
    -\sum_{ijkl} K_{ijkl} \big[(\mmdot ij)(\mmdot kl)+(\mmdot il)(\mmdot jk)-(\mmdot ik)(\mmdot jl)\big]
\end{align}
\end{widetext}
including Heisenberg exchange, DMI, uniaxial magnetic anisotropy, Zeeman interaction as well as biquadratic, 3-site and 4-site HOI.
The interaction strengths are given by $J_{ij}$, $\mathbf{D}_{ij}$, $K_{u}$, $\mathbf{B}_\text{ext}$, $B_{ij}$, $Y_{ijk}$ and $K_{ijkl}$ respectively.
The vector $\mathbf{D}_{ij}$ is aligned perpendicular to the connection line of the sites $i$ and $j$ within the plane of the magnetic film. Together with the isotropic $J_{ij}$ the shell resolved interactions strengths have been included for up to 11 shells of neighbors.
The magnetic anisotropy is treated in uniaxial approximation with the easy axis perpendicular to the film surface ($\mathbf{\hat{e}}_z$).
The external magnetic field is denoted by the component perpendicular to the film surface $\mathbf{B}_\text{ext}=B_\perp\mathbf{\hat{e}}_z$.

The summation of the HOI (Eq.~(\ref{eq:hamiltonian_HO_general})) covers double counting of indices $i,j,k,l$, respectively leading to three distinguished terms -- the biquadratic, 3-site and 4-site exchange -- restricted to shortest distances between magnetic moments on the hexagonal lattice, which is why we refer to the constants as $B_1$, $Y_1$, $K_1$ in the following, i.e.~using the same notation as in Ref.~\cite{paul2020}.
For a plaquette of four magnetic moments $\mathbf{m}_i, \mathbf{m}_j, \mathbf{m}_k, \mathbf{m}_l$ these interactions are illustrated in Fig.~\ref{fig:concept_barrier_vs_curvature}a. 
Here, a straight line between two magnetic moments $\mathbf{m}_i$ and $\mathbf{m}_j$ symbolizes a direct dot-product $\mathbf{m}_i\cdot\mathbf{m}_j$, while a curved connection between two of such pairs represents a product of these pairs.
See for example the top sketch for the 4-site HOI in Fig.~\ref{fig:concept_barrier_vs_curvature}b representing the term $(\mathbf{m}_i\cdot\mathbf{m}_j)(\mathbf{m}_l\cdot\mathbf{m}_k)$ and refer also to Ref.~\cite{hoffmann2020}.
General fourth-order HOI (Eq.~(\ref{eq:hamiltonian_HO_general})) include up to four different atomic sites $i$,$j$,$k$,$l$.
Thus, any specific HOI term can be uniquely identified through an array of six distances between the four sites:
\begin{align}
    d = [d_{ij}, d_{ik}, d_{il}, d_{jk}, d_{jl}, d_{kl} ]~,
    \label{eq:HOI_distances}
\end{align}
where the distances are encoded in units of the $n$-th nearest neighbor distance. 
The unique identifier of the 4-site HOI term is for example given by $d=[1,1,1,1,2,1]$ meaning that all sites are in nearest neighbor distance to each other except the sites $j$ and $l$, which are in second nearest neighbor distance (cf. Fig.~\ref{fig:concept_barrier_vs_curvature}b).
We implemented an efficient and general pre-processing routine automatically finding the contributing sites for a given HOI identifier (see ``Methods``).
The identifiers for the biquadratic, 3-site and 4-site HOI discussed in this work contain mostly nearest neighbor distances. However, the implementations are valid for any combination and distance of four magnetic moments and apply to any lattice model, even three-dimensional ones~\cite{beyer2025}.

To quantify the role of HOI, we compare two models previously used in Ref.~\cite{paul2020}: one including HOI and one neglecting HOI.
The interaction parameters for the latter model were obtained by DFT total energy calculations conducted in Ref.~\cite{dupe2014} for the ultrathin film system
\fccIr and in Ref.~\cite{haldar2018} for fcc-Pd/Fe/Rh(111). 
The HOI parameters ($B_1$, $Y_1$, $K_1$) were calculated in Ref.~\cite{paul2020}.
Note, that the extension of HOI modifies the first three Heisenberg exchange parameters $J_1, J_2$ and $J_3$. 

Above the critical magnetic field, $B_C$, the ferromagnetic (FM) alignment of magnetic moments corresponds to the ground state of the system.
For \fccIr these values are $B_C=2.75$~T~\cite{paul2020} for the model considering HOI explicitly and $B_C=3.17$~T~\cite{malottki2017} for the model neglecting HOI (see Supplementary Note~2 for values of fcc-Pd/Fe/Rh(111)).
We focus on the field regime $B_\perp\geq B_C$, where isolated skyrmions and antiskyrmions are meta-stable with respect to the FM state.
Such a skyrmion (antiskyrmion) at a field of $B_\perp-B_C=3.95$~T is visualized in Fig.~\ref{fig:concept_barrier_vs_curvature}a (inset of Fig.~\ref{fig:concept_barrier_vs_curvature}d).

\noindent{\textbf{Higher order exchange interactions in lifetime calculations.}}
Harmonic transition-state theory (HTST) provides a framework for definite calculations of average lifetime (Eq.~(\ref{eq:arrhenius})) of meta-stable magnetic configurations subject to thermal fluctuations. 
The Hamiltonian given in Eq.~(\ref{eq:hamiltonian}) defines the material-specific energy surface for magnetic textures -- such as skyrmions, antiskyrmions, and the FM state -- 
and allows for the computation of minimum energy paths (MEP) between them.
Such a path exhibits a first-order saddle point (SP), which is the configuration with the highest energy along the MEP and yields the energy barrier for a particular transition:
\begin{align}
    \Delta E = E_{\text{SP}}-E_\text{min}
    \label{eq:energybarrier}
\end{align}
such as the annihilation of a skyrmion or antiskyrmion into the FM state.
Fig.~\ref{fig:concept_barrier_vs_curvature}c (Fig.~\ref{fig:concept_barrier_vs_curvature}d) shows the total energy -- composed of all terms of Eq.~(\ref{eq:hamiltonian}) -- along the MEP of the skyrmion (antiskyrmion) annihilation at $B_\perp-B_C=3.95$~T when explicitly considering HOI.
The contribution of the biquadratic, 3-site and 4-site HOI to the total energy along the MEP is also shown in Fig.~\ref{fig:concept_barrier_vs_curvature}c,d.
It can be seen that the 4-site HOI term plays a prominent role in increasing the barrier for both skyrmions and antiskyrmions.

Until now, studies on the role of HOI in the meta-stability of skyrmions and antiskyrmions have focused on their effect on the energy barrier alone~\cite{paul2020, arya2025new} while neglecting entropic stabilization or destabilization effects included in the pre-exponential factor of the Arrhenius law (Eq.~(\ref{eq:arrhenius})).
This term reads~\cite{malottki2019,varentcova2020, goerzen2024thesis,malottki2025}:
\begin{align}
    \label{eq:preexp_factor}
    \nu_0 = \frac{\lambda}{\sqrt{2\pi\beta}}\mathrm{e}^{\Delta \mathcal{S} / k_{\text{B}}}~, 
\end{align}
where $\lambda$ describes the dynamical contribution~\cite{varentcova2020}, and $\Delta \mathcal{S} = \mathcal{S}^{\text{SP}}- \mathcal{S}^{\text{min}}$ is the relevant entropy difference between the transition state and the minimum. 
The $2N$ degrees of freedom of the system with $N$ magnetic moments are treated within harmonic approximation except the unstable mode at the SP and the contribution of possible zero modes:
\begin{equation}\label{eq:entropy}
     \frac{\Delta \mathcal{S}}{k_{\text{B}}} = \ln\left[ \frac{V^{\text{SP}}}{V^{\text{min}}} \left(\frac{2\pi}{\beta}\right)^{\frac{\Delta Z}{2}}\frac{\prod\limits_{k=1+Z^{\text{min}}}^{2N}\sqrt{\epsilon_k^{\text{min}}}}{\prod\limits_{k=2+Z^{\text{SP}}}^{2N}\sqrt{\epsilon_k^{\text{SP}}}}\right]~, 
\end{equation}
where $\epsilon_k^{\text{min/SP}}$ are the eigenvalues of the Hessian of the energy reflecting the curvature of the energy surface at the critical points.
An unequal number of zero modes at the minimum ($Z^{\text{min}}$) and the SP ($Z^{\text{SP}}$), with volumes $V^{\text{min}}$ and $V^{\text{SP}}$, yields an explicit temperature dependence with $\Delta Z= Z^{\text{min}}-Z^{\text{SP}}-1$ for the pre-exponential factor of skyrmions and antiskyrmions (see ``Methods``).
Note, that the true entropy difference $\Delta\mathcal{S}'$, which arises from free energy $\mathcal{F}$ of a canonical ensemble by $\mathcal{S}'=-\partial\mathcal{F}/\partial T$, relates to the numerically relevant entropy by $\Delta \mathcal{S}' = \Delta \mathcal{S} + k_{\text{B}}\Delta Z/2$.

The effect of the curvatures $\epsilon_k^{\text{min/SP}}$ on the lifetime of a metastable state is schematically illustrated in Fig.~\ref{fig:concept_barrier_vs_curvature}d, where two-dimensional energy landscapes are shown with identical energy differences between a minimum and an SP, but different shape of the energy surface around them. 
A flat minimum combined with a sharply curved energy landscape at the SP corresponds to a  large entropy barrier and yields a small $\nu_0$, leading to a longer lifetime in such a situation. 
Conversely, a sharply curved energy landscape at the minimum and a less curved energy landscape at the SP increase $\nu_0$, reducing the lifetime.
Despite its importance, the influence of HOI on the curvature of the energy landscape --
and thus on entropy -- has not yet been systematically explored.

Similar to bilinear exchange being mediated by isotropic tensors $\underline{\mathrm{J}}_{ij}\in\mathbb{R}^{3\times3}$ the HOI in Eq.~(\ref{eq:hamiltonian_HO_general}) can be expressed by isotropic tensors of fourth order $\underline{\mathrm{C}}_{ijkl}\in\mathbb{R}^{3\times3\times3\times3}$, which correspond to an interaction between lattice sites with indices $i,j,k,l$.
Due to the multi-linearity of HOI the calculation of the second-order derivatives with respect to the components of the magnetic moments comes down to sums of quadratic forms of $\underline{\mathrm{C}}_{ijkl}$ (see ``Methods``).
Fig.~\ref{fig:concept_barrier_vs_curvature}f schematically shows a part of the matrix of second-order derivatives associated with the four sites $i$, $j$, $k$ and $l$ depicted in Fig.~\ref{fig:concept_barrier_vs_curvature}b.
Each block in Fig.~\ref{fig:concept_barrier_vs_curvature}f corresponds to a $(3\times 3)$-block for the derivatives $\nabla_{ij}E_{\text{4-spin}}$ with respect to the $m_x$, $m_y$ and $m_z$-components of the magnetic moments, respectively.
The block colors indicate the contributing HOI terms in Fig.~\ref{fig:concept_barrier_vs_curvature}b.
Note, that only the contributions from the interactions presented in the mid row of Fig.~\ref{fig:concept_barrier_vs_curvature}b are shown, which already yields a complex pattern of contributions to the displayed sub-matrix. In comparison, bilinear interactions produce a relatively simple off-diagonal pattern of the matrix of second derivatives.

\begin{figure*}
    \centering
    \includegraphics{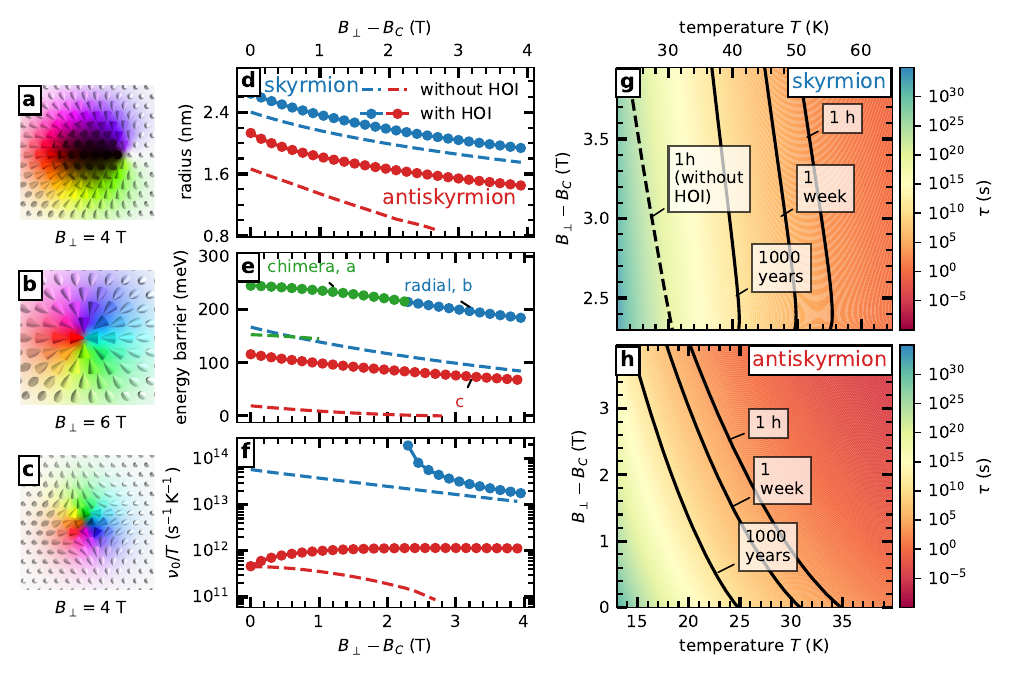}
    \caption{\textbf{Effect of higher-order exchange interactions on skyrmion and antiskyrmion lifetime.} \textbf{a,b,c} Saddle point configuration for the collapse of the isolated skyrmion at $B_\perp-B_C=1.25$~T~(\textbf{a}) and $B_\perp-B_C=3.25$~T~~(\textbf{b}) and antiskyrmion at $B_\perp-B_C=3.25$~T (\textbf{c}). \textbf{d} Radius of skyrmions (blue) and antiskyrmions (red) in \fccIr with (solid lines and circles) and without (dashed line) explicitly considering HOI as a function of the magnetic field. The magnetic field is given relative to the field $B_C$ for the onset of the field polarized phase for the respective energy model.
    \textbf{e} Energy barriers associated with the skyrmion and antiskyrmion collapse as a function of the magnetic field with and without (see Ref.~\cite{malottki2019,paul2020}) explicitly considering HOI. The field regime where the Chimera collapse of the skyrmion is favoured is colored green. 
    \textbf{f} Pre-exponential factor $\nu_0$ in units of temperature $T$ plotted over $B_\perp-B_C$ for skyrmions (blue) in the regime of the radial collapse and antiskyrmions (red) for the energy model considering HOI (circles) and not explicitly taking HOI into account (dashed lines) 
    \textbf{g} Mean lifetime $\tau$ (see colorbar) for different magnetic fields and different temperatures for the regime of radial skyrmion collapse. As a reference, the contour line for a lifetime of $1$~h for skyrmions without considering HOI explicitly is also given (see Ref.~\cite{malottki2019}). \textbf{h} Mean lifetime of antiskyrmions taking HOI into account.
    }
    \label{fig:lifetime}
\end{figure*}

\noindent{\textbf{Lifetime of skyrmions and antiskyrmions.}}
In this section we discuss numerically obtained observables, including the radius (Fig.~\ref{fig:lifetime}d), the energy barrier (Fig.~\ref{fig:lifetime}e), the pre-exponential factor (Fig.~\ref{fig:lifetime}f) and the lifetimes (Fig.~\ref{fig:lifetime}g,h) of skyrmions and antiskyrmions in Pd/Fe/Ir(111) as a function of the magnetic field for two cases: with and without explicit HOI. 
The radius -- obtained through a fit of a theoretical profile~\cite{bogdanov1994} to the magnetic texture -- increases upon the explicit inclusion of HOI (Fig.~\ref{fig:lifetime}d) but still the skyrmions and antiskyrmions are of the size of few nanometers agreeing well with experimental data~\cite{romming2013,hagemeister2015,muckel2021}.

For the energy barrier, pre-exponential factor, and lifetimes, identifying the relevant first-order SP -- characteristic of the respective collapse mechanism -- is crucial.
We compute SPs for annihilations of metastable states for various magnetic fields in the FM phase ($B_\perp>B_C$) using the GMMF method~\cite{schrautzer2025} (see ``Methods``).
When HOI are taken into account explicitly, the skyrmion annihilation proceeds via a Chimera collapse~\cite{meyer2019,muckel2021} for magnetic fields satisfying $B_\perp-B_C<2.25$~T.
The corresponding SP at a field of $B_\perp-B_C=1.25$~T is shown in Fig.~\ref{fig:lifetime}a.
At higher magnetic fields, the more commonly observed radial collapse mechanism~\cite{malottki2017,desplat2018,muller2018,bessarab2018} occurs, as illustrated in Fig.~\ref{fig:lifetime}b for $B_\perp-B_C=3.25$~T.
This SP can be characterized by four magnetic moments forming a Bloch point-like defect.

Within the explored parameter range, the collapse mechanism of the antiskyrmion corresponds to a radial shrinking of the configuration from the energy minimum (Fig.~\ref{fig:lifetime}c).
The energy barriers for the collapse of the skyrmions and antiskyrmions are shown in Fig.~\ref{fig:lifetime}e.
For comparison, the corresponding barriers obtained in the model neglecting HOI are also included.
In agreement with Ref.~\cite{paul2020}, explicitly considering HOI increases the energy barriers by approximately $100$~meV for fcc-Pd/Fe/Ir(111).

To quantify the thermal stability of skyrmions and antiskyrmions at a given temperature $T$ the mean-lifetime $\tau$ is computed using Eq.~(\ref{eq:arrhenius})~\cite{bessarab2012,bessarab2018} (see ``Methods``).
In magnetic systems, the entropic contribution -- contained in the pre-exponential factor $\nu_0$ -- can vary by several orders of magnitude as a function of the magnetic field~\cite{wild2017,malottki2019}.
Thus, it is essential to consider this factor explicitly.
Fig.~\ref{fig:lifetime}f shows the pre-exponential factor divided by the temperature as a function of the magnetic field $B_\perp-B_C$. 
Explicitly including HOI yields pre-exponential factors slightly increased but of the same order of magnitude for skyrmions, and approximately one order of magnitude larger for antiskyrmions, compared to the model neglecting HOI. This consequently has to be related to effects on the curvature of the energy surface, as discussed in the next section.
It should be noted that the pre-exponential factor is only shown in the regime $B_\perp-B_C\geq 2.25$~T, where the radial collapse mechanism of the skyrmion is favored (Fig.~\ref{fig:lifetime}f).

Using HTST considering both the effect of the energy barrier and the entropic contributions contained in the pre-exponential factor in Eq.~(\ref{eq:arrhenius}), the mean lifetime of skyrmions (Fig.~\ref{fig:lifetime}g) and antiskyrmions (Fig.~\ref{fig:lifetime}h) is evaluated across a range of temperatures and for varying $B_\perp-B_C$.
Comparison with the iso-lifetime line corresponding to one-hour in the model where we neglect HOI demonstrates that the explicit consideration of HOI leads to substantially increased thermal stability for both skyrmions and antiskyrmions.
While skyrmions remain stable for at least one hour at $T=50$~K within the considered field range when HOI are included, the temperature must be reduced to $T<30$~K to achieve comparable stability in the model neglecting HOI.
Furthermore, while the antiskyrmions are only stable up to fields of $B_\perp-B_C\approx 2.7$~T without HOI, explicit consideration of HOI yields antiskyrmion lifetimes of at least one hour for temperatures $T\leq20$~K in the field range of $B_\perp-B_C\approx 3$~T to $B_\perp-B_C\approx 4$~T. Thus, we record a beneficial effect of HOI on the lifetime and robustness of metastable magnetic states in this material.
We have performed similar calculations for \fccRh leading to qualitatively similar results presented in Supplementary Note~2.

\begin{figure*}
    \centering
    \includegraphics[width=1.0\linewidth]{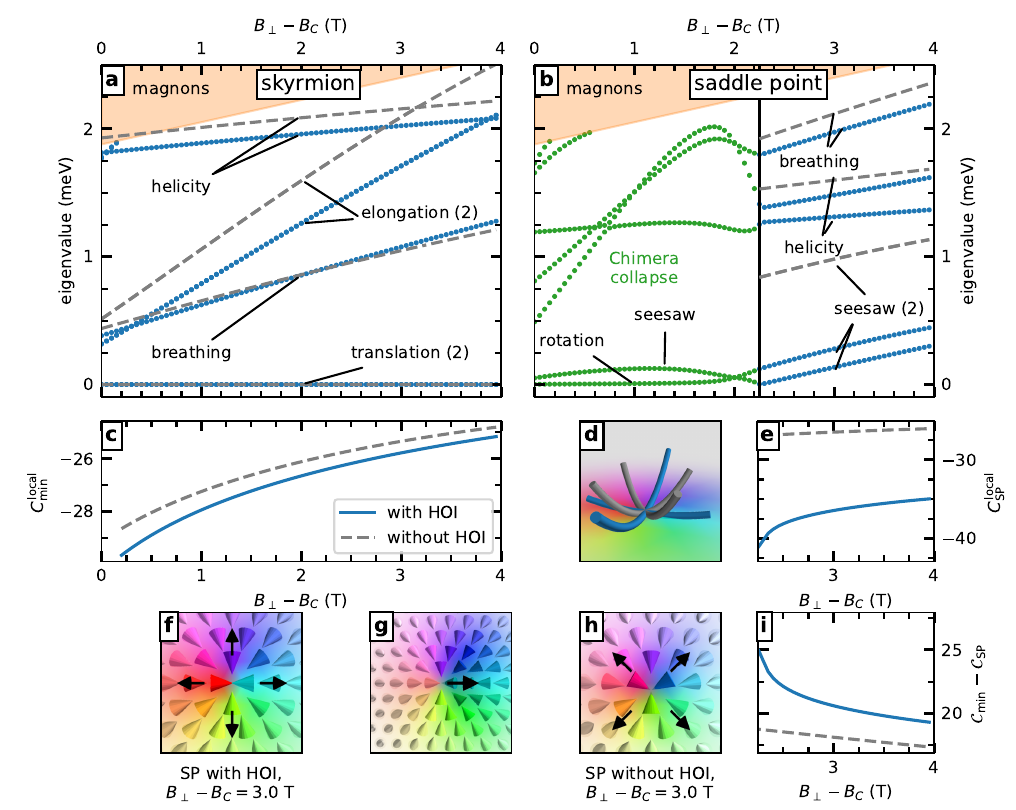}
    \caption{\textbf{HOI and curvature of the energy surface for the skyrmion.} \textbf{a,b} Local part of the eigenvalue spectrum of the Hessian matrix of the skyrmion (\textbf{a}, blue) and the skyrmion saddle point (SP) (\textbf{b}, green and blue) in \fccIr including HOI as a function of the magnetic field $B_\perp-B_C$. Eigenvalues above the magnon gap are indicated by the orange painted area. For reference the corresponding local eigenvalue spectra in the model without considering HOI explicitly are shown as gray dashed lines. Selected eigenmodes are labeled while the number in brakets gives their multiplicity (see surrounding text). \textbf{c,e} Curvature associated with the local eigenvalue spectrum (see Eq.~\ref{eq:curvature})) for the skyrmion (blue, \textbf{c}) and the corresponding SP (green and blue, \textbf{e}) for the energy model including HOI (solid line). The local curvatures of the SP for the model neglecting HOI are drawn with dashed gray lines. In \textbf{e} only the regime of the radial collapse is shown. \textbf{d} Energy along the seesaw eigenmodes of the SP including HOI (blue, SP in \textbf{f}) and of the SP excluding HOI (gray, SP in \textbf{h}). \textbf{g} Configuration obtained after displacing the SP in \textbf{f} along the indicated direction (black arrow) using the corresponding eigenvector of the seesaw mode. \textbf{i} Total curvature difference of the minimum and the SP for the regime of the radial collapse mechanism.}
    \label{fig:eigenvalues_sk}
\end{figure*}

\noindent{\textbf{Curvature of the energy surface.}}
The curvature of the energy surface in the vicinity of stationary points determines the entropic contributions to the lifetime of meta-stable configurations (Eq.~(\ref{eq:arrhenius})) as sketched in Fig.~\ref{fig:concept_barrier_vs_curvature}d.
Here, we explain the increased pre-exponential factor $\nu_0$ when HOI are explicitly included (cf.~Fig.~\ref{fig:lifetime}f), compared to the case in which HOI are neglected.
Figs.~\ref{fig:eigenvalues_sk}a,b present the eigenvalues of the Hessian of the energy (see Eq.~(\ref{eq:hessian})) as functions of the magnetic field $B_{\perp}-B_{C}$ for the skyrmion~(Fig.~\ref{fig:eigenvalues_sk}a) and for the SP (Fig.~\ref{fig:eigenvalues_sk}b) associated with skyrmion collapse in the \fccIr system.
For each stationary point, we divide the eigenspectrum into eigenmodes with eigenvalues below and above the boundary of the magnon gap.
For the Hamiltonian in Eq.~(\ref{eq:hamiltonian}) this gap reads: 
\begin{align}
    \epsilon_\text{mag}=\mu B_\perp +2K_u~.
    \label{eq:magnongap}
\end{align}
Eigenvalues below $\epsilon_{\text{mag}}$ correspond to excitations of the localized magnetic texture in the FM background and are referred to as local part of the eigenspectrum.
For the skyrmion, this local spectrum includes two zero modes (translations), a breathing mode, two degenerate elongation modes, and a helicity mode (see "Methods") 
The eigenvalue spectrum of the skyrmion including HOI closely resembles that of the skyrmion in the model without HOI~\cite{malottki2019}, depicted as dashed gray lines in Fig.~\ref{fig:eigenvalues_sk}a.
Quantitatively, the eigenvalues of the rotation and elongation modes of the skyrmion shift to lower values when HOI are explicitly included.

Considering HOI explicitly favors the Chimera collapse as the dominant skyrmion annihilation mechanism for $B-B_C\leq 2.25$~T, as shown in Fig.~\ref{fig:lifetime}e. 
The Chimera SP features a rotation mode (a zero mode) and a low-curvature seesaw mode.
This name was chosen since following this mode leads to a $180^\circ$-rotated Chimera SP via a configuration which is very similar to the radial SP and in fact constitutes a second-order SP on the energy landscape. 
Since HTST requires a clear separation between first- and second-order SPs, calculating the prefactor in this regime is technically challenging.
We present results for the pre-exponential factor of the Chimera collapse for the case of an applied in-plane magnetic field in Supplementary Note~3, since the main focus of this paper is the effect of HOI on lifetimes and the Chimera collapse has been discussed elsewhere~\cite{muckel2021}.

However, the seesaw mode is also relevant for the radial collapse regime. At $B_\perp-B_C=2.25$~T, its eigenvalue approaches zero, signaling a flat energy landscape between the Chimera and radial SPs, reminiscent of a second-order phase transition~\cite{schrautzer2022}. For higher fields, the radial SP becomes a first-order SP with two non-degenerate seesaw eigenvalues (Fig.~\ref{fig:eigenvalues_sk}b).
This SP (Fig.~\ref{fig:eigenvalues_sk}f) can be characterized by four magnetic moments, placed at the four vertices of a diamond, pointing toward each other.
The seesaw eigenmodes act as a deformation of the diamond-like SP configuration (Fig.~\ref{fig:eigenvalues_sk}f) into asymmetric states along the diagonals of the diamond (Fig.~\ref{fig:eigenvalues_sk}g). 
Since eigenvalues are directly associated with the curvature of the energy landscape the situation can be visualized by the parabolic energy dependence along the eigenvectors of the two seesaw eigenmodes which is shown in Fig.~\ref{fig:eigenvalues_sk}d in blue.

In contrast, neglecting HOI yields degenerate and larger seesaw eigenvalues (Fig.~\ref{fig:eigenvalues_sk}b, dashed), consistent with the triangular SP configuration (Fig.~\ref{fig:eigenvalues_sk}h) that preserves lattice symmetry and explains the degeneracy (Fig.~\ref{fig:eigenvalues_sk}h, gray).
Overall, explicit HOI reduce the curvature of the seesaw modes, and also lower the helicity and breathing eigenvalues of the SP associated to the radial collapse, similar to the shifts observed for the skyrmion elongation and helicity modes upon consideration of explicit HOI.

We define a measure $\mathcal{C}$ for curvature of the energy landscape at the minimum ($\mathcal{C}=\mathcal{C}_\text{min}$) and the SP ($\mathcal{C}=\mathcal{C}_\text{SP}$) excluding zero-modes, which are identified below the threshold $\delta_\text{zero}=10^{-12}$~eV, and the unstable mode of the SP with $\epsilon_1^{\text{SP}}<0$. Further, in order to identify thermodynamically relevant parts of the spectrum, we distinguish into local and magnon contribution, separated by $\epsilon_{\text{mag}}$ from Eq.~(\ref{eq:magnongap}):
\begin{align}
    \mathcal{C} &= \mathcal{C}^{\text{local}} + \mathcal{C}^{\text{magnon}}\\
    \mathcal{C} &=\sum\limits_{\delta_\text{zero}<\epsilon_k<\epsilon_\text{mag}}\ln\frac{\epsilon_k}{\epsilon_0} +\sum\limits_{\epsilon_k\geq\epsilon_\text{mag}}\ln\frac{\epsilon_k}{\epsilon_0}~,
    \label{eq:curvature}
\end{align}
where an energy scale $\epsilon_0=1$~eV was introduced to ensure dimensionless expressions in the logarithms.
This establishes an intuitive connection between curvatures and the entropy from Eq.~(\ref{eq:entropy}), which explicitly reads:
\begin{equation}\label{eq:pref_curvature}
     \frac{\Delta \mathcal{S}}{k_{\text{B}}} = \frac{\mathcal{C}_{\text{min}}-\mathcal{C}_\text{SP}}{2}
      + \ln\left[\frac{V^{\text{SP}}}{V^{\text{min}}} \left(\frac{2\pi}{\beta \epsilon_0}\right)^{\frac{\Delta Z}{2}}\right] ~.
\end{equation}
As shown in Fig.~\ref{fig:eigenvalues_sk}c,e, the curvature of the local part of the spectrum at the minimum is nearly unaffected by HOI, whereas the local curvature at the SP is strongly reduced due to shift of the seesaw eigenmodes. Since SP curvature enters Eq.~(\ref{eq:pref_curvature}) with a negative sign, this reduction, caused by HOI, enhances the annihilation rate. 
Including contributions above the magnon gap (Fig.~\ref{fig:eigenvalues_sk}i) reproduces the field dependence of the pre-exponential factor (Fig.~\ref{fig:lifetime}f).
However, it can be seen that the difference between both models is smaller in Fig.~\ref{fig:eigenvalues_sk}i than in Fig.~\ref{fig:eigenvalues_sk}e. 
This implies that the curvature contribution of the eigenspectrum above the magnon gap counteracts -- but not compensates -- the entropic destabilization by the eigenvalues of the local spectrum that are reduced upon explicit consideration of HOI.
In Supplementary Note~4 we present a similar discussion for the curvature of the energy surface for the antiskyrmion collapse.

\begin{figure*}
    \centering
    \includegraphics[width=1.0\linewidth]{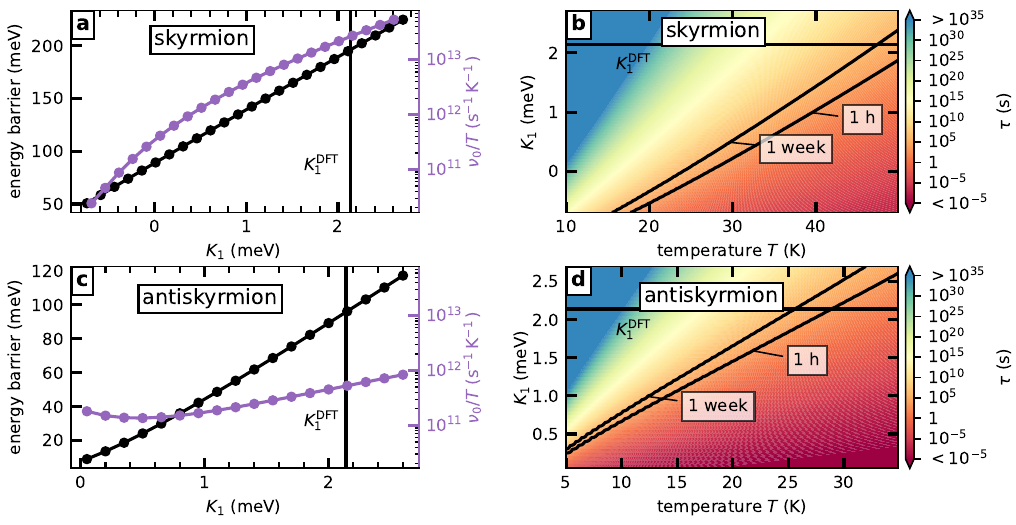}
    \caption{\textbf{Lifetimes of skyrmions and antiskyrmions as a function of the 4-site HOI.} \textbf{a},\textbf{b} Energy barrier (black) and pre-exponential factor $\nu_0/T$ (purple) for the skyrmion (\textbf{a}) and the antiskyrmion (\textbf{b}) annihilation in \fccIr for a magnetic field of $B_\perp=6$~T (skyrmions) and $B_\perp=4$~T (antiskyrmions) as a function of the 4-site HOI $K_1$. The value $K_1^{\text{DFT}}$ corresponds to the one determined in DFT calculations~\cite{paul2020}. \textbf{b},\textbf{d} Mean lifetime $\tau$ for isolated skyrmions (\textbf{b}) and antiskyrmions (\textbf{d}) for various temperatures $T$ and values of $K_1$ calculated within HTST. The one-week- and one-hour-isoline of $\tau$ are given as black lines.}
    \label{fig:lifetime_varyK}
\end{figure*}

\noindent{\textbf{Tuning the 4-site HOI interaction.}}
Now we turn to the influence of variations in the 4-site HOI parameter $K_1$ (see Eq.~(\ref{eq:hamiltonian})) on the lifetimes of skyrmions (Fig.~\ref{fig:lifetime_varyK}b) and antiskyrmions (Fig.~\ref{fig:lifetime_varyK}d) with respect to thermally activated annihilation into the FM state.
Since spin spiral energies are degenerate with respect to the 4-site HOI parameter, the parametrization of the bilinear interaction constants obtained
in DFT remains valid even if $K_1$ is varied \cite{paul2020}.
This allows us to analyze the role of the 4-site HOI in stabilizing topological magnetic textures.
In Ref.~\cite{paul2020}, a linear dependence of the energy barriers on $K_1$ was reported for both skyrmions and antiskyrmions.
Our calculations reproduce this behavior (Fig.~\ref{fig:lifetime_varyK}a,c).
While skyrmions remain metastable even for negative $K_1$, the energy barrier for antiskyrmions vanishes as $K_1\to 0$~meV.

Here, we also compute the pre-exponential factor for skyrmions and antiskyrmions as a function of $K_1$. 
We find that the pre-exponential factor for antiskyrmions (Fig.~\ref{fig:lifetime_varyK}c) remains nearly constant across the considered range, whereas for skyrmions it decreases by about three orders of magnitude upon reducing $K_1$ (Fig.~\ref{fig:lifetime_varyK}a).
The change of pre-exponential factor and energy barrier with strength of the 4-site HOI, leads to opposite effects on the thermal stability of skyrmions. While the increase of  the energy barrier with $K_1$ also leads to an enhanced lifetime, the increase of the pre-exponential factor results in a decrease of the lifetime as seen from the Arrhenius law (cf.~Eq.~(\ref{eq:arrhenius})).
Note that the variation of $K_1$ within the considered range $K_1\in[-0.75,2.7]$~meV represents realistic values for the 4-site HOI in transition-metal thin film systems \cite{paul2020,gutzeit2021,gutzeit2022,beyer2025}.

The lifetime of skyrmions (Fig.~\ref{fig:lifetime_varyK}b) and antiskyrmions (Fig.~\ref{fig:lifetime_varyK}d) changes drastically upon variation of $K_1$.
For instance, at $T=40$~K the predicted lifetime of a skyrmion with $K_1=1.0$~meV is about $45$~minutes.
Increasing $K_1$ by only $0.5$~meV extends the lifetime to roughly $18$~days. The effect of HOI on the stability is also visible from the increase of the temperature for a given lifetime.
For example, a mean lifetime of above one hour is obtained below a temperature of about 30~K for $K_1=0$ and increases to above 40~K for a small 4-site four spin term of $K_1=1$~meV.
For antiskyrmions, we observe a similar trend, however, at overall shorter lifetimes at a given temperature due to the lower energy barriers (Fig.~\ref{fig:lifetime_varyK}c).

\begin{figure}
    \centering
    \includegraphics{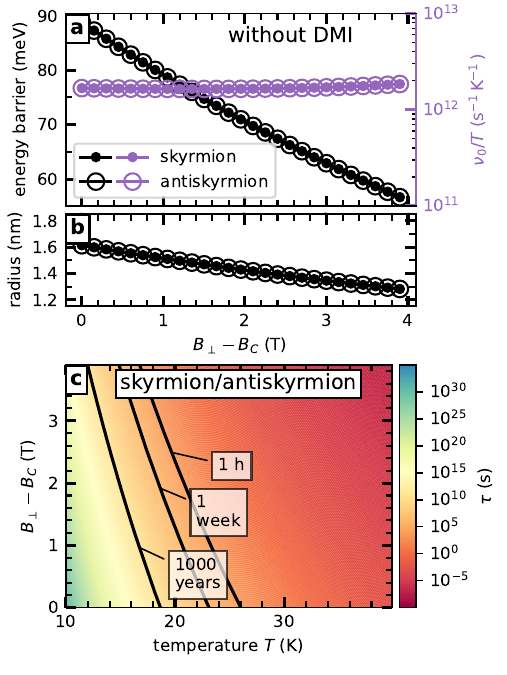}
    \caption{\textbf{Skyrmion and antiskyrmion lifetime including HOI and neglecting DMI.}
    \textbf{a} Energy barriers (black) and pre-exponential factors $\nu_0$ (purple) of skyrmions (filled circles) and antiskyrmions (open circles) in \fccIr considering HOI and vanishing DMI for varying the magnetic field relative to the critical field $B_\perp-B_C$. \textbf{b} Radius of skyrmions and antiskyrmions as a function of the magnetic field. \textbf{c}: Mean lifetime $\tau$ (see colorbar) of the above mentioned isolated magnetic textures as a function of the temperature $T$ and the external magnetic field $B_{\perp}-B_C$.}
    \label{fig:lifetime_zero_DMI}
\end{figure}

\noindent{\textbf{Skyrmions and antiskyrmions without DMI}.}
Designing systems hosting isolated skyrmions with maximum thermal stability typically requires large DMI constants and/or exchange frustration. 
In Ref.~\cite{paul2020}, it was predicted that both skyrmions and antiskyrmions are metastable when HOI are explicitly included, even in the absence of DMI, based on calculations of the energy barriers for their collapse into the FM state.
However, thermal stability cannot be deduced from the barrier height alone.
In this section, we present the calculated pre-exponential factor in the Arrhenius law and the corresponding mean lifetime for fcc-Pd/Fe/Ir(111), including HOI but neglecting DMI (see Fig.~\ref{fig:lifetime_zero_DMI}).
The isolated skyrmion and its topological counterpart, the antiskyrmion, exhibit significant and identical energy barriers for their collapse into the FM state (Fig.~\ref{fig:lifetime_zero_DMI}a) consistent with previous work~\cite{paul2020}.
The origin of the energy barrier is nearly entirely due to the effect of the 4-site HOI while the exchange frustration contributes only little.

In addition, we quantify the pre-exponential factors and demonstrate that they are also degenerate when DMI is neglected (Fig.~\ref{fig:lifetime_zero_DMI}a).
The values obtained for the pre-exponential factor are basically independent of the applied magnetic field and of a similar order of magnitude as previously reported for skyrmions in ultrathin films including DMI~\cite{malottki2019}.
This independence can be related to the skyrmion and antiskyrmion radius (Fig.~\ref{fig:lifetime_zero_DMI}b), which is only changing little upon variation of the applied field.
Neglecting DMI leads to a smaller values of the radius of skyrmions and antiskyrmions and to a weaker field-dependence of the size compared to the radius of these textures including DMI (cf.~Fig.~\ref{fig:lifetime}d).
Furthermore, the SP configurations corresponding to the annihilation of these textures into the FM state consist of only few magnetic moments forming the Bloch point-like defect and consequently their size also does not change upon variation of the applied field.
Since the size and the shape of the configurations of the minima and the SPs change only little, the curvature values of the energy landscape at the stationary points, expressed by the respective eigenvalues of the Hessian, and consequently also the pre-exponential factors show a minor dependence to the applied magnetic field.
The corresponding eigenvalue spectra for the skyrmion and antiskyrmion -- again degenerate -- and the SP when neglecting DMI but explicitly considering HOI are presented in Supplementary Note~5.

Since the pre-exponential factors are the same for skyrmions and antiskyrmions, their mean lifetimes -- as obtained from HTST -- are also identical (Fig.~\ref{fig:lifetime_zero_DMI}c).
Note, that in the case of skyrmions in a Hamiltonian without DMI the helicity mode of the minimum and the SP configuration becomes a zero mode~\cite{lin2016,goerzen2023} (``Methods").
The results in Fig.~\ref{fig:lifetime_zero_DMI}c indicate that skyrmions and antiskyrmions remain metastable over a broad range of magnetic fields $B_{\perp}-B_C$, at timescales and temperatures relevant to 
low temperature STM experiments, even in the absence of DMI.

\section*{}\label{sec:discussion}
\noindent{\large{\textbf{Discussion}}}\par
In this work, we have investigated the effect of higher-order exchange interactions (HOI) on the lifetimes of isolated skyrmions and antiskyrmions in \fccIr and fcc-Pd/Fe/Rh(111), focusing on their thermally activated annihilation into the FM state.
By implementing general fourth-order HOI terms for the energy, gradient and Hessian of the system into our atomistic spin simulations code, we are able to provide a comprehensive, quantitative description of these lifetimes within the framework of HTST.
This enables a quantitative description of lifetimes in systems with HOI that goes well beyond simple energy-barrier arguments for the stability of skyrmions and antiskyrmions.
Since HOI can play a key role in many materials such as transition-metal interfaces~\cite{heinze2011,yoshida2012,paul2020,Kroenlein2018,Spethmann2020,gutzeit2022,beyer2025} and 2D van der Waals magnets and heterostructures \cite{kartsev2020,Ni2021,xu2022,li2023,arya2025new} our results represent an important step towards realistic stability predictions for topological magnetic textures in these systems.

The enhanced energy barriers observed upon explicitly considering HOI lead to markedly increased lifetimes of skyrmions and antiskyrmion (see Fig.~\ref{fig:lifetime}e,g,h), which has important implications for the design of robust topological magnetic bits.
Furthermore, we report a slight reduction of the stability of skyrmions and antiskyrmions due to changed entropic contributions when including HOI in materials that are subject to our investigation (Fig~\ref{fig:lifetime}f).
We were able to identify the origin of the increased pre-exponential factor for the skyrmion collapse as a reduced curvature of the energy landscape due to the local Hessian eigenvalue spectrum at the SP (see Fig.~\ref{fig:eigenvalues_sk}c,e,i).
The dynamic contributions to the pre-exponential factor change only little upon considering HOI explicitly.

For both cases -- the skyrmion and the antiskyrmion collapse -- the HOI terms affect the curvature in the direction of the localized excitations at the SP significantly more than the corresponding curvatures at the minimum. 
These SP configurations exhibit larger canting angles between adjacent magnetic moments than the minima.
Since the contribution of the HOI terms to the energy is strongly affected by strong canting of the magnetization this is likely to be the origin of the pronounced effect of HOI on the energy landscape at the SP. 
This makes entropic stabilization by HOI, due to increased (decreased) curvature of the energy surface at the SP (minimum), a promising route to engineer skyrmion and antiskyrmion stability. 
A particularly striking result is the sensitivity of lifetimes to the 4-site interaction parameter $K_1$, which underlines the critical role of HOI in determining skyrmion and antiskyrmion stability. 
It also suggests that targeted manipulation of $K_1$, for example by material choice or interface engineering, could provide a powerful strategy to design skyrmions with lifetimes tailored for specific experimental or technological applications.
Remarkably, our study reveals a strong dependence of the pre-exponential factor on the strength of the 4-site interaction for skyrmions.
This suggests that accurate lifetime predictions require full entropic modeling rather than assuming a constant prefactor.

Finally, we also predict that both skyrmions and antiskyrmions remain metastable in the complete absence of DMI -- stabilized almost entirely by HOI -- at timescales and temperatures relevant to experiments. 
While switching off the DMI in \fccIr represents a toy model, our results highlight the potential of HOI as a promising route for designing systems that host topological magnetic textures in materials with inversion symmetry or without heavy transition-metals required for large spin-orbit coupling. 
These findings are timely, given the recent progress in 2D van der Waals magnets where DMI is often weak or even absent, positioning HOI as a promising stabilization mechanism in future material systems.

\section*{}\label{sec:methods}
\noindent{\large{\textbf{Methods}}}\par
\noindent{\textbf{Calculation of the pre-exponential factor.}}
The mean-lifetime $\tau$ for thermally activated annihilation of skyrmions and antiskyrmions into the FM state are computed in this work using the Arrhenius law~(Eq.~(\ref{eq:arrhenius})) within the framework of harmonic transition state theory (HTST)~\cite{bessarab2012}. 
In HTST the transition state for a particular reaction mechanism described by an MEP connecting two minima (skyrmion/antiskyrmion and FM state) is chosen as the hyperplane through a first order SP perpendicular to its unstable mode. 
This SP is the configuration of highest energy along the MEP.
The contributions to the Arrhenius law are, on the one hand, the energy barrier $\Delta E$, which is given as the energy difference of the SP and the energy minimum configuration (see Eq.~(\ref{eq:energybarrier})).
On the other hand, the pre-exponential factor $\nu_0$ is given by the entropic and the dynamic contribution $\lambda$ to the transition~\cite{varentcova2020} (Eq.~(\ref{eq:preexp_factor})).

Note, that while several formulations exist for the pre-exponential factor~\cite{bessarab2012,desplat2018,malottki2019}, Eq.~(\ref{eq:preexp_factor}) is particularly useful to discuss entropic effects (Eq.~(\ref{eq:entropy})).
While almost all degrees of freedom are treated in harmonic approximation, translations, rotations or modes changing the helicity of the localized magnetic texture are often better treated in Goldstone mode approximation~\cite{bessarab2018,malottki2019,goerzen2023} and their contribution to the entropy then corresponds to the zero mode volume $V$.
Additionally, a factor $\rho$ for the multiplicity of identical SP per unit cell~\cite{malottki2019} is required.
In this work, it is either $\rho=1$ or $\rho=2$.
Supplementary Note~6 presents the values of $\rho$ of the pre-exponential factor calculations of this work.
Note, that the helicity degree of freedom of the skyrmion in the HO model neglecting DMI, for which the pre-exponential factor is shown in Fig.~\ref{fig:lifetime_zero_DMI}a, is also a zero mode with volume \cite{goerzen2023}
\begin{equation}
    V_{\text{hel}} = 2\pi\sum_{n=1}^N\left[1 - \left(m_n^z\right)^2\right] ~.
\end{equation}
The respective degree of freedom describes the continuous deformation between Néel- and Bloch-type skyrmions, which comes down to a collective rotation of all spins in the $m_x-m_y$ plane.

\noindent{\textbf{Identification of first order saddle points.}}
It is essential to identify the corresponding SP in order
to calculate the energy barrier and the pre-exponential factor of the skyrmion and antiskyrmion collapse into the FM state.
In this work, these SPs are determined by combining the geodesic nudged elastic band method (GNEB)~\cite{bessarab2015} with the recently developed geodesic minimum mode following method (GMMF)~\cite{schrautzer2025}. The GNEB method was used to determine the skyrmion or antiskyrmion collapse mechanism and the SP for the maximum magnetic field considered for the specific parameter set ($B_\perp-B_C=4$~T). Subsequently, the GMMF method was initialized using this SP and the magnetic field was reduced by a small amount of $\Delta B=0.05$~T. 

In contrast to the GNEB method, in which the GNEB force for a series of replicas of the system (path) is minimized, the GMMF method operates on a single magnetic configuration. The GMMF force corresponds to the negative energy gradient where the direction of the gradient is inverted along the direction of the minimum mode of the Hessian of the energy:
\begin{align}
    \mathbf{F}_{\text{GMMF}}=-\mathbf{\nabla}E-2(\mathbf{\nabla}E\cdot \mathbf{v}_1)\mathbf{v}_1~.
    \label{eq:GMMF_force}
\end{align}
This yields a minimization of the energy along all degrees of freedom except one, while the energy is maximized along the direction of the eigenvector $\mathbf{v}_1$ of the minimum mode. If the GMMF method is initialized with a configuration in the vicinity of an SP, it converges to a first order SP within a few iterations. The computational bottleneck is the repeated calculation of the eigenvector of the lowest eigenvalue of the Hessian matrix of the energy of the system.

Recently, we introduced Rayleigh Quotient minimization~\cite{schrautzer2025}, a method that makes an efficient calculation of the minimum mode feasible and is based on finite difference calculations of the energy gradient without the need for the explicit calculation of the Hessian matrix. The efficient combination of these methods allowed us to calculate the activation energies with a high resolution with respect to the values of the magnetic field with almost negligible computational effort.

\noindent{\textbf{Hessian matrix including HOI.}}
This section describes the calculation of the Hessian matrix for general fourth-order HOI, which is required to determine the pre-exponential factors of the skyrmion and antiskyrmion collapse. 
We approximate the energy surface $E(\mathbf{m})$ on the $2N$-dimensional configuration space $\mathbf{m}\in\bigotimes_n^N\mathbb{S}^2$ up to second order in the vicinity of critical points $\mathbf{m}_0$.
Since first order terms vanish in these points the series expansions in terms of tangential deflections $\delta\mathbf{m}$ reads
\begin{equation}\label{eq:harmonic_approx_energy}
    E(\mathbf{m}_0, \delta\mathbf{m}) = E(\mathbf{m}_0) + \frac{1}{2}\delta\mathbf{m}^{\text{T}}\mathrm{H}\delta\mathbf{m} + \mathcal{O}(\delta\mathbf{m}^3)~.
\end{equation}
Due to the constraint of magnetic moments having fixed length the Hessian $\mathrm{H}\in\mathbb{R}^{2N\times2N}$ is constrained to the tangent space of the magnetic configuration:
\begin{equation}
    \mathrm{H}_{ij} = \mathrm{P}_{i} (\nabla_{ij}\mathcal{L})\mathrm{P}_{j}^T ~,\quad \mathcal{L} = E -\sum_{i=1}^N (\mathbf{m}_i^2-1)\omega_i
    \label{eq:hessian}
\end{equation}
with the Lagrange function $\mathcal{L}$ and Lagrange multipliers $\omega_i=\mathbf{m}\cdot\nabla_iE / 2$. Here $\nabla_i = \frac{\partial}{\partial\mathbf{m}_i}$ is the partial derivative with respect to the coordinates of the magnetic moment $\mathbf{m}_i$ and $\mathrm{P}_i\in\mathbb{R}^{2\times3}$ is the projection on the tangent space of $\mathbf{m}_i$~\cite{muller2018,varentcova2020}.
In our setup derivatives of the Lagrange multiplier $\omega_i$ can be omitted, because they describe changes of the energy under variation of the spin length. 
Thus, they are perpendicular to the manifold associated with configuration space and vanish in the quadratic form of Eq.~(\ref{eq:harmonic_approx_energy}), since they are perpendicular to tangential deflections $\delta\mathbf{m}$.

While bilinear interactions between magnetic moments have been considered extensively in the framework of HTST~\cite{bessarab2018,desplat2018,malottki2019,varentcova2020,goerzen2023}, the novelty of this work is the consideration of HOI (cf.~Eq.~(\ref{eq:hamiltonian_HO_general})). 
Gradients with respect to coordinates of magnetic moments are computed as
\begin{equation}\label{eq:HO_gradient}
    \nabla_{i}E_{\text{4-spin}} = -4\sum_{jkl}C_{ijkl}\left( \mathbf{m}_{k}\cdot\mathbf{m}_{l}\right)\mathbf{m}_{j} ~.
\end{equation}
Similar to bilinear exchange $E_{\text{exc}}\propto\mathbf{m}_i^{\text{T}}\underline{\mathrm{J}}_{ij}\mathbf{m}_j$ being mediated by a tensor $\underline{\mathrm{J}}_{ij}\in\mathbb{R}^{3\times3}$, HOI are mediated by a tensor $\underline{\mathrm{C}}_{ijkl}\in\mathbb{R}^{3\times3\times3\times3}$. Since both the bilinear exchange as well as HOI are invariant under global rotation of all magnetic moments, and therefore isotropic in configuration space, the HOI tensor can be expressed in the basis of isotropic tensors of fourth-order, which are usually denoted as products of Kronecker $\delta$ symbols
\begin{equation}
    \underline{\mathrm{C}}_{ijkl} = C_{ijkl}\delta_{ij}\delta_{kl}
\end{equation}
with interaction constant $C_{ijkl}$.
It is commonly known that this basis contains only two more linearly independent elements, $\delta_{ik}\delta_{jl}$ and $\delta_{il}\delta_{kj}$, which in our case correspond to certain permutations of indices $(i,j,k,l)$.
In matrix form these tensors can be constructed as
\begin{equation}
    \label{eq:4spin_tensors}
    \begin{split}
    \delta_{ij}\delta_{kl} &= \left[\begin{matrix}\left[\begin{matrix}1 & 0 & 0\\0 & 1 & 0\\0 & 0 & 1\end{matrix}\right] & \left[\begin{matrix}0 & 0 & 0\\0 & 0 & 0\\0 & 0 & 0\end{matrix}\right] & \left[\begin{matrix}0 & 0 & 0\\0 & 0 & 0\\0 & 0 & 0\end{matrix}\right]\\\left[\begin{matrix}0 & 0 & 0\\0 & 0 & 0\\0 & 0 & 0\end{matrix}\right] & \left[\begin{matrix}1 & 0 & 0\\0 & 1 & 0\\0 & 0 & 1\end{matrix}\right] & \left[\begin{matrix}0 & 0 & 0\\0 & 0 & 0\\0 & 0 & 0\end{matrix}\right]\\\left[\begin{matrix}0 & 0 & 0\\0 & 0 & 0\\0 & 0 & 0\end{matrix}\right] & \left[\begin{matrix}0 & 0 & 0\\0 & 0 & 0\\0 & 0 & 0\end{matrix}\right] & \left[\begin{matrix}1 & 0 & 0\\0 & 1 & 0\\0 & 0 & 1\end{matrix}\right]\end{matrix}\right] \\
    \delta_{ik}\delta_{jl} &= \left[\begin{matrix}\left[\begin{matrix}1 & 0 & 0\\0 & 0 & 0\\0 & 0 & 0\end{matrix}\right] & \left[\begin{matrix}0 & 1 & 0\\0 & 0 & 0\\0 & 0 & 0\end{matrix}\right] & \left[\begin{matrix}0 & 0 & 1\\0 & 0 & 0\\0 & 0 & 0\end{matrix}\right]\\\left[\begin{matrix}0 & 0 & 0\\1 & 0 & 0\\0 & 0 & 0\end{matrix}\right] & \left[\begin{matrix}0 & 0 & 0\\0 & 1 & 0\\0 & 0 & 0\end{matrix}\right] & \left[\begin{matrix}0 & 0 & 0\\0 & 0 & 1\\0 & 0 & 0\end{matrix}\right]\\\left[\begin{matrix}0 & 0 & 0\\0 & 0 & 0\\1 & 0 & 0\end{matrix}\right] & \left[\begin{matrix}0 & 0 & 0\\0 & 0 & 0\\0 & 1 & 0\end{matrix}\right] & \left[\begin{matrix}0 & 0 & 0\\0 & 0 & 0\\0 & 0 & 1\end{matrix}\right]\end{matrix}\right] \\
    \delta_{il}\delta_{kj} &= \left[\begin{matrix}\left[\begin{matrix}1 & 0 & 0\\0 & 0 & 0\\0 & 0 & 0\end{matrix}\right] & \left[\begin{matrix}0 & 0 & 0\\1 & 0 & 0\\0 & 0 & 0\end{matrix}\right] & \left[\begin{matrix}0 & 0 & 0\\0 & 0 & 0\\1 & 0 & 0\end{matrix}\right]\\\left[\begin{matrix}0 & 1 & 0\\0 & 0 & 0\\0 & 0 & 0\end{matrix}\right] & \left[\begin{matrix}0 & 0 & 0\\0 & 1 & 0\\0 & 0 & 0\end{matrix}\right] & \left[\begin{matrix}0 & 0 & 0\\0 & 0 & 0\\0 & 1 & 0\end{matrix}\right]\\\left[\begin{matrix}0 & 0 & 1\\0 & 0 & 0\\0 & 0 & 0\end{matrix}\right] & \left[\begin{matrix}0 & 0 & 0\\0 & 0 & 1\\0 & 0 & 0\end{matrix}\right] & \left[\begin{matrix}0 & 0 & 0\\0 & 0 & 0\\0 & 0 & 1\end{matrix}\right]\end{matrix}\right]
    \end{split}
\end{equation}
In this notation the indices $i,j$ address the position of each $3\times3$-matrix within the tensors above.
The indices $k,j$ address the position of $1$ or $0$ within these submatrices.
Note, that other representations of this basis exist as well.
Intuitively, the linearly independent basis elements reflect that some permutations of lattice sites in fourth-order HOI lead to different energies
\begin{equation}
\begin{split}
    (\mathbf{m}_i\cdot\mathbf{m}_j)(\mathbf{m}_k\cdot\mathbf{m}_l) &\neq (\mathbf{m}_i\cdot\mathbf{m}_k)(\mathbf{m}_j\cdot\mathbf{m}_l) \\ &\neq (\mathbf{m}_i\cdot\mathbf{m}_l)(\mathbf{m}_k\cdot\mathbf{m}_j)
\end{split}
\end{equation}
while linear dependent permutations leave the energy invariant
\begin{equation}
\begin{split}
    (\mathbf{m}_i\cdot\mathbf{m}_j)(\mathbf{m}_k\cdot\mathbf{m}_l) &= (\mathbf{m}_k\cdot\mathbf{m}_l)(\mathbf{m}_i\cdot\mathbf{m}_j) \\ &= (\mathbf{m}_l\cdot\mathbf{m}_k)(\mathbf{m}_j\cdot\mathbf{m}_i)~.
\end{split}
\end{equation}
In order to find the most general form of second derivatives of the fourth-order HOI energy with respect to $\mathbf{m}_i$ and $\mathbf{m}_j$ all linearly independent permutations have to be considered
\begin{align}
\nabla_{ij} E_{\text{4-spin}} 
  &= \sum_{kl} \mathbf{m}_k^{\text{T}} \big(
        C_{ijkl} \delta_{ij} \delta_{kl}
      + C_{ikjl} \delta_{ik} \delta_{jl} \nonumber \\[-7pt]
  &\qquad\qquad\quad
      + C_{ilkj} \delta_{il} \delta_{kj} 
    \big) \mathbf{m}_l 
    \label{eq:hessian_4spintensor_basis_1} \\
  &= \sum_{kl} \big(
        C_{ijkl} \, \mathbf{m}_k \odot \mathbf{m}_l
      + C_{ikjl} \, \mathbf{m}_l \otimes \mathbf{m}_k \nonumber \\[-7pt]
  &\qquad\qquad\qquad\qquad
      + C_{ilkj} \, \mathbf{m}_k \otimes \mathbf{m}_l
    \big)
    \label{eq:hessian_4spintensor_basis_2}
\end{align}
where we used the abbreviations
\begin{align}
    \mathbf{m}_i \otimes \mathbf{m}_j &= \left(\begin{array}{ccc}
         m_i^xm_j^x & m_i^xm_j^y & m_i^xm_j^z \\
         m_i^ym_j^x & m_i^ym_j^y & m_i^ym_j^z \\
         m_i^zm_j^x & m_i^zm_j^y & m_i^zm_j^z 
    \end{array}\right) \\
     \mathbf{m}_i \odot \mathbf{m}_j &= \left(\begin{array}{ccc}
         \mathbf{m}_i\cdot\mathbf{m}_j & 0 & 0 \\
         0 & \mathbf{m}_i\cdot\mathbf{m}_j & 0 \\
         0 & 0 & \mathbf{m}_i\cdot\mathbf{m}_j
    \end{array}\right)
\end{align}
The step between Eq.~(\ref{eq:hessian_4spintensor_basis_1}) and Eq.~(\ref{eq:hessian_4spintensor_basis_2}) involves quadratic forms of spins with every $3\times3$-submatrix denoted within the tensors of Eq.~(\ref{eq:4spin_tensors}) \cite{goerzen2024thesis}.
Since the last two terms in Eq.~(\ref{eq:hessian_4spintensor_basis_2}) are related by $\mathbf{m}_k\otimes\mathbf{m}_l = (\mathbf{m}_l\otimes\mathbf{m}_k)^T$ (transposing inverts order of magnetic moments) the construction of Hessian blocks comes down to the two cases:
\begin{equation}\label{eq:HO_hessian}
        \nabla_{ij}E_{\text{4-spin}} = -4\sum_{kl} \left\{\begin{array}{cl}
                C_{ijkl}\ \mathbf{m}_k\odot\mathbf{m}_l & \text{for}~ (i,j,k,l)\\
                C_{ikjl}\ \mathbf{m}_l\otimes\mathbf{m}_k & \text{for}~ (i,k,j,l)
            \end{array}\right.~,
\end{equation}
where only the order of indices has to be considered, including variations with repeating indices which address 3-site and biquadratic HOI terms.
\begin{figure}
    \centering
    \includegraphics[width=1.0\linewidth]{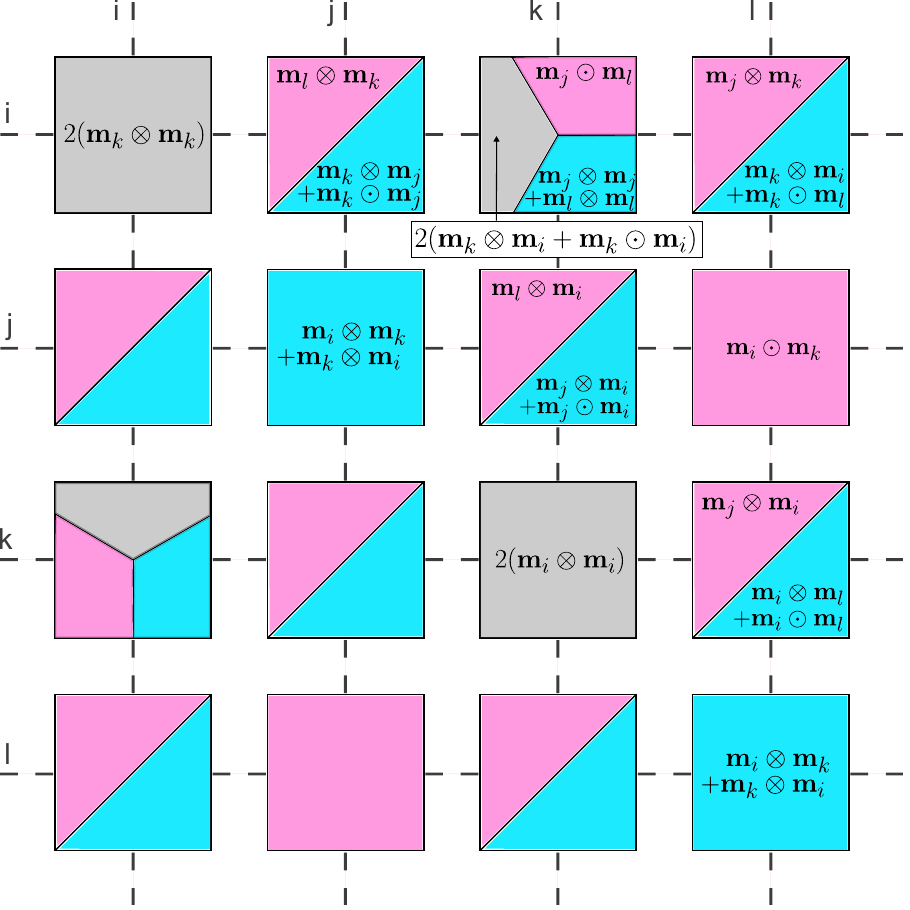}
    \caption{Schematic adjacency pattern for sub-matrix of the Hessian for four magnetic moments ($\mathbf{m}_i$, $\mathbf{m}_j$, $\mathbf{m}_k$, $\mathbf{m}_l$) in nearest neighbor distance of each other (see Fig.~\ref{fig:concept_barrier_vs_curvature}a). Each block $ij$ corresponds to the $3\times 3$ cartesian components of the second derivatives of the energy in Eq.~(\ref{eq:hamiltonian_HO_general}) with respect to the magnetic moments. The blocks colors refer to the specific HOI term contributing (biquadratic (gray), 3-site (cyan) and 4-site (pink) HOI.}
    \label{fig:method_hessian_contr}
\end{figure}
Fig.~\ref{fig:method_hessian_contr} shows exemplarily contributions of HOI to the Hessian blocks of four magnetic moments within nearest neighbor distance to each other.
It is noteworthy that $\nabla_{ij}E_{\text{4-spin}} = \left(\nabla_{ji}E_{\text{4-spin}}\right)^T$, so that the Hessian itself is symmetric, as required on torsion-free configuration spaces.

\begin{figure*}
    \centering
    \includegraphics[width=1.0\linewidth]{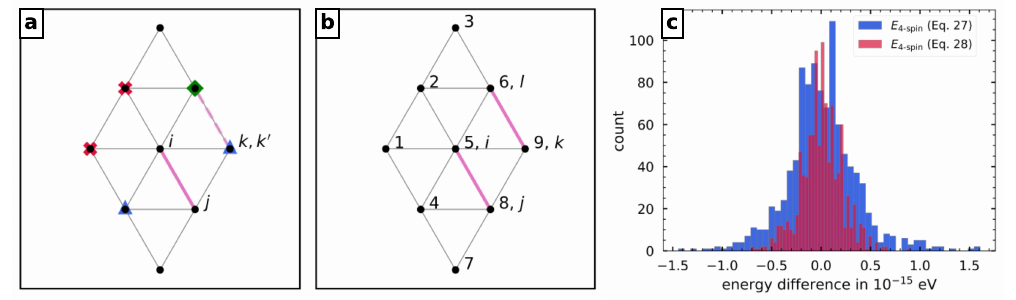}
    \caption{
    \textbf{Pairfinder illustrations and energy derivative consistency test.}
    \textbf{a,b} Schematic illustration of the Pairfinder algorithm utilized to calculate the neighbor relations for a specific HOI term for a 3 by 3 segment of the hexagonal lattice. Exemplary the derivation of the 4-site HOI term is shown. The moments $i$ and $j$ participate in the first scalar product and $k$, $l$ in the latter. The sites marked by blue triangles satisfy the adjacency to $i$ and $j$ during the generation of the neighbor $k, k'$. For $k, k'$ on the right only the site marked by the green diamond satisfies the adjacency of the fourth moment $l$ to the others.
    \textbf{b} The final set of four moments with pink lines as the scalar products is illustrated in addition to the indices of the atomic sites.
    \textbf{c} Histogram of via Euler's theorem numerically calculated 4-spin energy differences within the atomistic spin simulation code 
    \textsc{Spinaker}. The difference between the energy and the energy via the gradient in Eq.~(\ref{eq:test_impl_HO_findiff1}) (red) and the energy via the matrix of second-order derivatives in Eq.~(\ref{eq:test_impl_HO_findiff2}) (blue) of $1000$ states with random magnetization is compared.
    } 
    \label{fig:pairfinder_hist}
\end{figure*}
\noindent{\textbf{Implementation of the HOI.}}
General fourth-order HOI (Eq.~(\ref{eq:hamiltonian_HO_general})) were implemented in our atomistic simulation package \textsc{Spinaker} for arbitrary lattice models, including three-dimensional ones.
The code is capable of an efficient computation of the energy, gradients and Hessians of user-input fourth-order HOI over arbitrary distances of the lattice.
Each term contributing to the general form of the fourth-order HOI, e.g. the biquadratic, 3-site or 4-site term, can be uniquely identified through an array of six distances (Eq.~(\ref{eq:HOI_distances})).
These distances are given in units of $n$-th nearest neighbor distances between the atomic sites of the respective lattice model.
Based on this adjacency information the neighbor relations of a specific HOI are derived in a pre-processing step through the "Pairfinder"-algorithm explained below.
A pseudo-code representation of the core of the Pairfinder algorithm is given in Alg. \ref{alg:parfinder}.

The lattice of atoms is constructed by repetitions of the chemical unit cell containing one or more atomic species.
Consider a central atomic site with index $i$. 
Figs.~\ref{fig:pairfinder_hist}a,b illustrate small 3 by 3 segments of a hexagonal lattice with an exemplary central atomic site $i$ and an integer index assigned to each site of these nine unit cells (Fig.~\ref{fig:pairfinder_hist}b).
In Fig.~\ref{fig:pairfinder_hist}a the Pairfinder is illustrated for the 4-site HOI (denoted by the interaction strength $K$ in Eq.~(\ref{eq:hamiltonian}).
In this case, the array of distances is $[d_{ij},d_{ik},d_{il},d_{jk},d_{jl},d_{kl}]=[1,1,1,1,2,1]$, involving only nearest neighbors except for the connection of $j$ and $l$: $d_{jl}=2$.

The outermost loop in Alg.~\ref{alg:parfinder} iterates over the neighbors of the atomic site $i$ satisfying the constraint $d_{ij}$.
For the 4-site HOI $j$ corresponds to a site in nearest neighbor distance to $i$ and for the specific system in Fig.~\ref{fig:pairfinder_hist}b we have e.g. $j=8$ for $i=5$.
Subsequently the index shift $n_{ij}=j-i$ is stored, which is here $n_{ij}=3$.
Ascending one level in the nested loops in Alg.~\ref{alg:parfinder}, the next loop iterates over the neighbors with the distance constraint $d_{ik}$ and the index shift $n_{ik}$ is stored.
For the 4-site HOI term these are again the sites in nearest neighbor distance to $i$ ($d_{ik}=1$). 
Consider e.g. the atomic site with index $k=9$ to be selected at this stage (see Fig.~\ref{fig:pairfinder_hist}b) with $n_{ik}=4$.
Now it has to be checked whether the selected neighbors $j$ and $k$ of $i$ fulfill also the distance constraint $d_{jk}$, which is in this case again $d_{jk}=1$.
For this purpose the neighbors $k'$ in distance $d_{jk}$ of $j$ are iterated.
For each candidate $k'$ it is tested if the index shifts of the candidate sites correspond to a closed path via $n_{ij}+n_{jk'}-n_{ik}=0$.
If this equation is satisfied, as for the site $k'$ marked on the right in Fig.~\ref{fig:pairfinder_hist}a, we set $k=k'$ and continue. The sites marked with blue triangles are candidates for $k, k'$ satisfying the equation above. 
However, if this condition is not satisfied $k'$ is discarded and the loop over the neighbors of $j$ in distance $d_{jk}$ continues with the next candidate $k'$.
The next embedded loop iterates over the neighbors of $i$ within distance $d_{il}$ to the site $i$ to produce candidates for the fourth site $l$.
For the 4-site HOI these are again the nearest neighbors of $i$.
For each candidate $l$ we now have to test whether it also fulfills the distance constraints $d_{jl}$ and $d_{kl}$, which are $d_{jl}=2$ and $d_{kl}=1$ here.
Similar to the procedure above this is achieved by iterating the neighbors $l'$ of $j$ in distance $d_{jl}$ and iterating the neighbors $l''$ of $k$ in distance $d_{kl}$. 
For each $l'$ and $l''$ it is checked whether $n_{ij}+n_{jl'}-n_{il}=0$ and $n_{ik}+n_{kl''}-n_{il}=0$, respectively.
Only if both conditions are satisfied the candidate $l$ completes the set of the four magnetic moments searched for the specific HOI term and the six index shifts are stored.

Note, that for each specific HOI term several sets fulfilling the above conditions exist.
Consider again the 4-site HOI, where the site $l$ has to be in second nearest neighbor distance to $j$ and in nearest neighbor distance to $k$.
In Fig.~\ref{fig:pairfinder_hist}a this is only satisfied by the lattice site marked with the green diamond if $j$ and $k$ are chosen as schematically depicted in Fig.~\ref{fig:pairfinder_hist}.
However, also the lattice sites $j=4$ and $k=1$ satisfy the conditions for the first three magnetic moments participating in the 4-site HOI term.
This leaves the site $l=2$ as the only option to complete this set. 
Iterating over lattice sites corresponding to the array of distances recursively and evaluating conditions of closed paths ensures, finding all the sets of interacting lattice sites for a specific HOI, $12$ for the case of the 4-site HOI.

Given the translational symmetry of the lattice it is sufficient to calculate these relative index shifts only once for a representative atomic site $i$. 
Thus, the computational complexity of this pre-processing step is negligible with respect to the thousands of energy and energy gradient calculations employed in a typical simulation, e.g.~an iterative energy minimization algorithm.

Note, that the above representation uses single integer indices to address atomic sites and represent relative positions of sites for the sake of clarity.
However, in practice \textsc{Spinaker} stores multi-indices $(i_{\mathbf{a}_1},i_{\mathbf{a}_2},i_{\mathbf{a}_3},i_{\text{species}})$ addressing a site via the coefficients of the linear combination of lattice vectors to reach its unit cell together with a unique index of the atomic species within this cell.
This allows to detect neighbors connected via periodic boundaries. 
These are handled with a precomputed lookup mask storing for each site whether a neighbor is present or not, which allows also to simulate defects and missing atoms.

The Pairfinder algorithm explained above provides the necessary adjacency information used in the calculation of derivatives of the HOI energy with respect to the components of the magnetic moments -- gradient and the matrix of second-order derivatives (Eqs.~(\ref{eq:HO_gradient}),(\ref{eq:HO_hessian})).
On the one hand, the implementations are verified through comparing the results with analytical solutions, e.g. comparing the energy of magnetic structures with literature \cite{Kurz2000}.
On the other hand, the consistency between the energy calculation and derivation of the first- and second-order derivatives with respect to the magnetization is checked via Euler's theorem for homogeneous functions.
The fourth-order HOI term is a homogeneous function with order four of $3N$ variables $m_i^\alpha$ for $i=1,\dots,N$ and $\alpha\in\{x,y,z\}$:
\begin{align}
    E_\text{4-spin}(s\mathbf{m}_1,\dots,s\mathbf{m}_N)=s^4E_\text{4-spin}(\mathbf{m}_1,\dots,\mathbf{m}_N)
\end{align}
with $s\neq0$.
Differentiating both sides of this equation with respect to $s$ and taking the limit of the results with $s\to 1$ yields:
\begin{align}
    4E_\text{4-spin}=\sum\limits_i \mathbf{m}_i\cdot(\mathbf{\nabla}_iE_\text{4-spin})~.
     \label{eq:test_impl_HO_findiff1}
\end{align}
Therefore the energy is expressed in terms of the gradient.
Similarly, the energy can be restored from the matrix of second derivatives via
\begin{align}
    12E_\text{4-spin}=\sum_{i,j} \mvec i ^T \cdot (\nabla_{ij} E_{\text{4-spin}}) \cdot \mvec j~.
     \label{eq:test_impl_HO_findiff2}
\end{align}
Fig.~\ref{fig:pairfinder_hist}c shows the energy difference of the left sides of Eqs.~(\ref{eq:test_impl_HO_findiff1}) 
and (\ref{eq:test_impl_HO_findiff2}) computed with the HOI energy routine and the respective right sides, computed using the gradient and second-order derivative routines, respectively.
The fluctuations are on the order of machine precision.
The histogram for errors in the computation via the Hessian shows a broader distribution of numerical fluctuations, due to a quadratically increased amount of floating point operations.
\begin{figure}[!ht]
\begin{algorithm}[H]
    \caption{Pairfinder}
    \label{alg:parfinder}
    \begin{algorithmic}
        \LineComment declare $i$ as the central atomic site
        \LineComment reset the set-counter for the specific HOI term
        \State $N\gets 0$
        \LineComment iterates the neighbors $j$ of $i$
            \For{$j$ in neighbors($d_{ij}$)}
                \LineComment store the index shift between $i$ and $j$
                \State $n_{ij} \gets \text{indexshift($d_{ij}, j$)}$
                \For{$k$ in neighbors($d_{ik}$)}
                \State $n_{ik} \gets \text{indexshift($d_{ik}, k$)}$
                    \For{$k'$ in neighbors($d_{jk}$)}
                        \State $n_{jk'} \gets \text{indexshift($d_{jk}, k'$)}$
                        \If{$n_{ij}+n_{jk'}-n_{ik} \neq 0$}
                            \State cycle
                        \EndIf
                        \LineComment now $k=k'$, proceed with $l,l'$ and $l''$
                        \For{$l$ in neighbors($d_{il}$)}
                            \State $n_{il} \gets \text{indexshift($d_{il}, l$)}$
                            \For{$l'$ in neighbors($d_{jl}$)}
                                \State $n_{jl'} \gets \text{indexshift($d_{jl}, l'$)}$
                                \If{$n_{ij}+n_{jl'}-n_{il} \neq 0$}
                                    \State cycle
                                \EndIf
                                \For{$l''$ in neighbors($d_{kl}$)}
                                    \State $n_{kl''} \gets \text{indexshift($d_{kl}, l''$)}$
                                    \If{$n_{ik}+n_{kl''}-n_{il} \neq 0$}
                                        \State cycle
                                    \EndIf
                                    \LineComment set of 4 moments found
                                    \State N $\gets N+1$
                                    \LineComment store index shifts
                                    \State n[N] $\gets [n_{ij}, n_{ik}, n_{il}, n_{jk}, n_{jl}, n_{kl}]$
                                \EndFor
                            \EndFor
                        \EndFor
                    \EndFor
                \EndFor
            \EndFor
    \end{algorithmic}
\end{algorithm}
\end{figure}

\section*{}
\noindent{{\large\textbf{Data availability}.}\newline The data presented in this paper are available from the authors upon reasonable request.

\section*{}
\noindent{{\large\textbf{Code availability}.}\newline The atomistic spin simulation code is available from the authors upon reasonable request.

\section*{Acknowledgement}
H. S. acknowledges financial support from the Icelandic Research Fund (grant No. 239435). P F. B. acknowledges financial support from the Icelandic Research Fund (Grants No. 2410333 and No. 217750), the University of Iceland Research Fund (Grant No. 15673), the Swedish Research Council (Grant No. 2020-05110), and the Crafoord Foundation (Grant No. 20231063). S.~Ha and S.~He.~gratefully acknowledge financial support from the Deutsche Forschungsgemeinschaft (DFG, German Research Foundation) through SPP2137 ``Skyrmionics" (project no.~462602351) and project no.~418425860. This work was performed using HPC resources available at the Kiel University Computing Centre. H. S. thanks S. Paul and S. von Malottki for valuable discussions.

\section*{}
\noindent{\large{\textbf{Additional information}}}\newline
\noindent{\textbf{Supplementary Information}} accompanies this paper.

\noindent{\textbf{Competing interests:}} The authors declare that there are no non-financial or financial competing interests.

\section*{}
\noindent{\large{\textbf{Author contributions}}}\newline
H.S. performed the calculations and prepared the figures. B.B. implemented and tested higher-order interactions in the code supported by M.A.G. and H.S.
All authors discussed the results. H.S. and M.A.G. wrote the first version of the manuscript, and all authors contributed to the final version.

\section*{}
\noindent{\large{\textbf{References}}}\newline
\bibliographystyle{naturemag}
\bibliography{bib_submission}

@article{bogdanov1989,
  title={Thermodynamically stable “vortices” in magnetically ordered crystals. The mixed state of magnets},
  author={Bogdanov, Alexei N and Yablonskii, DA},
  journal={Zh. Eksp. Teor. Fiz},
  volume={95},
  number={1},
  pages={178},
  year={1989}
}

@Article{nayak2017,
author={Nayak, Ajaya K.
and Kumar, Vivek
and Ma, Tianping
and Werner, Peter
and Pippel, Eckhard
and Sahoo, Roshnee
and Damay, Franoise
and R{\"o}{\ss}ler, Ulrich K.
and Felser, Claudia
and Parkin, Stuart S. P.},
title={Magnetic antiskyrmions above room temperature in tetragonal Heusler materials},
journal={Nature},
year={2017},
month={Aug},
day={01},
volume={548},
number={7669},
pages={561-566},
issn={1476-4687},
doi={10.1038/nature23466}
}

@Article{nagaosa2013,
author={Nagaosa, Naoto
and Tokura, Yoshinori},
title={Topological properties and dynamics of magnetic skyrmions},
journal={Nature Nanotechnology},
year={2013},
month={Dec},
day={01},
volume={8},
number={12},
pages={899-911},
issn={1748-3395},
doi={10.1038/nnano.2013.243}
}

@Article{iwasaki2013,
author={Iwasaki, Junichi
and Mochizuki, Masahito
and Nagaosa, Naoto},
title={Universal current-velocity relation of skyrmion motion in chiral magnets},
journal={Nature Communications},
year={2013},
month={Feb},
day={12},
volume={4},
number={1},
pages={1463},
issn={2041-1723},
doi={10.1038/ncomms2442}
}

@Article{sampaio2013,
author={Sampaio, J.
and Cros, V.
and Rohart, S.
and Thiaville, A.
and Fert, A.},
title={Nucleation, stability and current-induced motion of isolated magnetic skyrmions in nanostructures},
journal={Nature Nanotechnology},
year={2013},
month={Nov},
day={01},
volume={8},
number={11},
pages={839-844},
issn={1748-3395},
doi={10.1038/nnano.2013.210}
}

@Article{zhou2014,
author={Zhou, Yan
and Ezawa, Motohiko},
title={A reversible conversion between a skyrmion and a domain-wall pair in a junction geometry},
journal={Nature Communications},
year={2014},
month={Aug},
day={13},
volume={5},
number={1},
pages={4652},
issn={2041-1723},
doi={10.1038/ncomms5652}
}

@Article{Woo2016,
author={Woo, Seonghoon
and Litzius, Kai
and Kr{\"u}ger, Benjamin
and Im, Mi-Young
and Caretta, Lucas
and Richter, Kornel
and Mann, Maxwell
and Krone, Andrea
and Reeve, Robert M.
and Weigand, Markus
and Agrawal, Parnika
and Lemesh, Ivan
and Mawass, Mohamad-Assaad
and Fischer, Peter
and Kl{\"a}ui, Mathias
and Beach, Geoffrey S. D.},
title={Observation of room-temperature magnetic skyrmions and their current-driven dynamics in ultrathin metallic ferromagnets},
journal={Nature Materials},
year={2016},
month={May},
day={01},
volume={15},
number={5},
pages={501-506},
issn={1476-4660},
doi={10.1038/nmat4593}
}

@Article{Fert2017,
author={Fert, Albert
and Reyren, Nicolas
and Cros, Vincent},
title={Magnetic skyrmions: advances in physics and potential applications},
journal={Nature Reviews Materials},
year={2017},
month={Jun},
day={13},
volume={2},
number={7},
pages={17031},
issn={2058-8437},
doi={10.1038/natrevmats.2017.31}
}

@Article{Fert2013,
author={Fert, Albert
and Cros, Vincent
and Sampaio, Jo{\~a}o},
title={Skyrmions on the track},
journal={Nature Nanotechnology},
year={2013},
month={Mar},
day={01},
volume={8},
number={3},
pages={152-156},
issn={1748-3395},
doi={10.1038/nnano.2013.29}
}

@article{pinna2018,
  title = {Skyrmion Gas Manipulation for Probabilistic Computing},
  author = {Pinna, D. and Abreu Araujo, F. and Kim, J.-V. and Cros, V. and Querlioz, D. and Bessiere, P. and Droulez, J. and Grollier, J.},
  journal = {Phys. Rev. Appl.},
  volume = {9},
  issue = {6},
  pages = {064018},
  numpages = {21},
  year = {2018},
  month = {Jun},
  publisher = {American Physical Society},
  doi = {10.1103/PhysRevApplied.9.064018}
}

@Article{song2020,
author={Song, Kyung Mee
and Jeong, Jae-Seung
and Pan, Biao
and Zhang, Xichao
and Xia, Jing
and Cha, Sunkyung
and Park, Tae-Eon
and Kim, Kwangsu
and Finizio, Simone
and Raabe, J{\"o}rg
and Chang, Joonyeon
and Zhou, Yan
and Zhao, Weisheng
and Kang, Wang
and Ju, Hyunsu
and Woo, Seonghoon},
title={Skyrmion-based artificial synapses for neuromorphic computing},
journal={Nature Electronics},
year={2020},
month={Mar},
day={01},
volume={3},
number={3},
pages={148-155},
issn={2520-1131},
doi={10.1038/s41928-020-0385-0}
}

@Article{grollier2020,
author={Grollier, J.
and Querlioz, D.
and Camsari, K. Y.
and Everschor-Sitte, K.
and Fukami, S.
and Stiles, M. D.},
title={Neuromorphic spintronics},
journal={Nature Electronics},
year={2020},
month={Jul},
day={01},
volume={3},
number={7},
pages={360-370},
issn={2520-1131},
doi={10.1038/s41928-019-0360-9}
}

@article{muehlbauer2009,
author = {S. Mühlbauer  and B. Binz  and F. Jonietz  and C. Pfleiderer  and A. Rosch  and A. Neubauer  and R. Georgii  and P. Böni },
title = {Skyrmion Lattice in a Chiral Magnet},
journal = {Science},
volume = {323},
number = {5916},
pages = {915-919},
year = {2009},
doi = {10.1126/science.1166767}}

@Article{yu2010,
author={Yu, X. Z.
and Onose, Y.
and Kanazawa, N.
and Park, J. H.
and Han, J. H.
and Matsui, Y.
and Nagaosa, N.
and Tokura, Y.},
title={Real-space observation of a two-dimensional skyrmion crystal},
journal={Nature},
year={2010},
month={Jun},
day={01},
volume={465},
number={7300},
pages={901-904},
issn={1476-4687},
doi={10.1038/nature09124}
}

@Article{heinze2011,
author={Heinze, Stefan
and von Bergmann, Kirsten
and Menzel, Matthias
and Brede, Jens
and Kubetzka, Andr{\'e}
and Wiesendanger, Roland
and Bihlmayer, Gustav
and Bl{\"u}gel, Stefan},
title={Spontaneous atomic-scale magnetic skyrmion lattice in two dimensions},
journal={Nature Physics},
year={2011},
month={Sep},
day={01},
volume={7},
number={9},
pages={713-718},
issn={1745-2481},
doi={10.1038/nphys2045}
}

@article{romming2013,
author = {Niklas Romming  and Christian Hanneken  and Matthias Menzel  and Jessica E. Bickel  and Boris Wolter  and Kirsten von Bergmann  and André Kubetzka  and Roland Wiesendanger },
title = {Writing and Deleting Single Magnetic Skyrmions},
journal = {Science},
volume = {341},
number = {6146},
pages = {636-639},
year = {2013},
doi = {10.1126/science.1240573}}

@Article{meyer2019,
author={Meyer, Sebastian
and Perini, Marco
and von Malottki, Stephan
and Kubetzka, Andr{\'e}
and Wiesendanger, Roland
and von Bergmann, Kirsten
and Heinze, Stefan},
title={Isolated zero field sub-10{\thinspace}nm skyrmions in ultrathin {Co} films},
journal={Nature Communications},
year={2019},
month={Aug},
day={23},
volume={10},
number={1},
pages={3823},
issn={2041-1723},
doi={10.1038/s41467-019-11831-4}
}

@Article{muckel2021,
author={Muckel, Florian
and von Malottki, Stephan
and Holl, Christian
and Pestka, Benjamin
and Pratzer, Marco
and Bessarab, Pavel F.
and Heinze, Stefan
and Morgenstern, Markus},
title={Experimental identification of two distinct skyrmion collapse mechanisms},
journal={Nature Physics},
year={2021},
month={Mar},
day={01},
volume={17},
number={3},
pages={395-402},
issn={1745-2481},
doi={10.1038/s41567-020-01101-2}
}

@Article{moreau2016,
author={Moreau-Luchaire, C.
and Moutafis, C.
and Reyren, N.
and Sampaio, J.
and Vaz, C. A. F.
and Van Horne, N.
and Bouzehouane, K.
and Garcia, K.
and Deranlot, C.
and Warnicke, P.
and Wohlh{\"u}ter, P.
and George, J.-M.
and Weigand, M.
and Raabe, J.
and Cros, V.
and Fert, A.},
title={Additive interfacial chiral interaction in multilayers for stabilization of small individual skyrmions at room temperature},
journal={Nature Nanotechnology},
year={2016},
month={May},
day={01},
volume={11},
number={5},
pages={444-448},
issn={1748-3395},
doi={10.1038/nnano.2015.313}
}

@Article{Boulle2016,
author={Boulle, Olivier
and Vogel, Jan
and Yang, Hongxin
and Pizzini, Stefania
and de Souza Chaves, Dayane
and Locatelli, Andrea
and Mente{\c{s}}, Tevfik Onur
and Sala, Alessandro
and Buda-Prejbeanu, Liliana D.
and Klein, Olivier
and Belmeguenai, Mohamed
and Roussign{\'e}, Yves
and Stashkevich, Andrey
and Ch{\'e}rif, Salim Mourad
and Aballe, Lucia
and Foerster, Michael
and Chshiev, Mairbek
and Auffret, St{\'e}phane
and Miron, Ioan Mihai
and Gaudin, Gilles},
title={Room-temperature chiral magnetic skyrmions in ultrathin magnetic nanostructures},
journal={Nature Nanotechnology},
year={2016},
month={May},
day={01},
volume={11},
number={5},
pages={449-454},
issn={1748-3395},
doi={10.1038/nnano.2015.315}
}

@Article{Soumyanarayanan2017,
author={Soumyanarayanan, Anjan
and Raju, M.
and Gonzalez Oyarce, A. L.
and Tan, Anthony K. C.
and Im, Mi-Young
and Petrovi{\'{c}}, A.  P.
and Ho, Pin
and Khoo, K. H.
and Tran, M.
and Gan, C. K.
and Ernult, F.
and Panagopoulos, C.},
title={Tunable room-temperature magnetic skyrmions in 
{Ir/Fe/Co/Pt} multilayers},
journal={Nature Materials},
year={2017},
month={Sep},
day={01},
volume={16},
number={9},
pages={898-904},
issn={1476-4660},
doi={10.1038/nmat4934}
}

@Article{han2019topological,
author={Han, Myung-Geun
and Garlow, Joseph A.
and Liu, Yu
and Zhang, Huiqin
and Li, Jun
and DiMarzio, Donald
and Knight, Mark W.
and Petrovic, Cedomir
and Jariwala, Deep
and Zhu, Yimei},
title={Topological Magnetic-Spin Textures in Two-Dimensional van {der Waals} {Cr$_2$Ge$_2$Te$_6$}},
journal={Nano Letters},
year={2019},
month={Nov},
day={13},
publisher={American Chemical Society},
volume={19},
number={11},
pages={7859-7865},
issn={1530-6984},
doi={10.1021/acs.nanolett.9b02849}
}

@Article{ding2019observation,
author={Ding, Bei
and Li, Zefang
and Xu, Guizhou
and Li, Hang
and Hou, Zhipeng
and Liu, Enke
and Xi, Xuekui
and Xu, Feng
and Yao, Yuan
and Wang, Wenhong},
title={Observation of Magnetic Skyrmion Bubbles in a van der {Waals} Ferromagnet {Fe$_3$GeTe$_2$}},
journal={Nano Letters},
year={2020},
month={Feb},
day={12},
publisher={American Chemical Society},
volume={20},
number={2},
pages={868-873},
issn={1530-6984},
doi={10.1021/acs.nanolett.9b03453}
}

@Article{wu2020neel,
author={Wu, Yingying
and Zhang, Senfu
and Zhang, Junwei
and Wang, Wei
and Zhu, Yang Lin
and Hu, Jin
and Yin, Gen
and Wong, Kin
and Fang, Chi
and Wan, Caihua
and Han, Xiufeng
and Shao, Qiming
and Taniguchi, Takashi
and Watanabe, Kenji
and Zang, Jiadong
and Mao, Zhiqiang
and Zhang, Xixiang
and Wang, Kang L.},
title={N{\'e}el-type skyrmion in {WTe$_2$/Fe$_3$GeTe$_2$} van der {W}aals heterostructure},
journal={Nature Communications},
year={2020},
month={Jul},
day={31},
volume={11},
number={1},
pages={3860},
issn={2041-1723},
doi={10.1038/s41467-020-17566-x}
}

@Article{yang2020creation,
author={Yang, M.
and Li, Q.
and Chopdekar, R. V.
and Dhall, R.
and Turner, J.
and Carlstr{\"o}m, J. D.
and Ophus, C.
and Klewe, C.
and Shafer, P.
and N'Diaye, A. T.
and Choi, J. W.
and Chen, G.
and Wu, Y. Z.
and Hwang, C.
and Wang, F.
and Qiu, Z. Q.},
title={Creation of skyrmions in van der {Waals} ferromagnet {Fe$_3$GeTe$_2$} on {(Co/Pd)$_n$} superlattice},
journal={Science Advances},
year={2020},
publisher={American Association for the Advancement of Science},
volume={6},
number={36},
pages={eabb5157},
doi={10.1126/sciadv.abb5157}
}

@article{wu2021van,
author = {Wu, Yingying and Francisco, Brian and Chen, Zhijie and Wang, Wei and Zhang, Yu and Wan, Caihua and Han, Xiufeng and Chi, Hang and Hou, Yasen and Lodesani, Alessandro and Yin, Gen and Liu, Kai and Cui, Yong-tao and Wang, Kang L. and Moodera, Jagadeesh S.},
title = {A Van der {Waals} Interface Hosting Two Groups of Magnetic Skyrmions},
journal = {Advanced Materials},
volume = {34},
number = {16},
pages = {2110583},
doi = {https://doi.org/10.1002/adma.202110583},
year = {2022}
}

@article{liu2024magnetic,
author = {Liu, Chen and Zhang, Senfu and Hao, Hongyuan and Algaidi, Hanin and Ma, Yinchang and Zhang, Xi-Xiang},
title = {Magnetic Skyrmions above Room Temperature in a van der {Waals} Ferromagnet {Fe$_3$GaTe$_2$}},
journal = {Advanced Materials},
volume = {36},
number = {18},
pages = {2311022},
doi = {https://doi.org/10.1002/adma.202311022},
year = {2024}
}

@Article{kartsev2020,
author={Kartsev, Alexey
and Augustin, Mathias
and Evans, Richard F. L.
and Novoselov, Kostya S.
and Santos, Elton J. G.},
title={Biquadratic exchange interactions in two-dimensional magnets},
journal={npj Computational Materials},
year={2020},
month={Oct},
day={09},
volume={6},
number={1},
pages={150},
issn={2057-3960},
doi={10.1038/s41524-020-00416-1}
}

@article{Ni2021,
  title = {Giant Biquadratic Exchange in {2D} Magnets and Its Role in Stabilizing Ferromagnetism of {${\mathrm{NiCl}}_{2}$} Monolayers},
  author = {Ni, J. Y. and Li, X. Y. and Amoroso, D. and He, X. and Feng, J. S. and Kan, E. J. and Picozzi, S. and Xiang, H. J.},
  journal = {Phys. Rev. Lett.},
  volume = {127},
  issue = {24},
  pages = {247204},
  numpages = {7},
  year = {2021},
  month = {Dec},
  publisher = {American Physical Society},
  doi = {10.1103/PhysRevLett.127.247204},
}

@article{xu2022,
author = {Xu, Changsong and Li, Xueyang and Chen, Peng and Zhang, Yun and Xiang, Hongjun and Bellaiche, Laurent},
title = {Assembling Diverse Skyrmionic Phases in {Fe$_3$GeTe$_2$} Monolayers},
journal = {Advanced Materials},
volume = {34},
number = {12},
pages = {2107779},
doi = {https://doi.org/10.1002/adma.202107779},
year = {2022}
}

@article{li2023,
  title = {Topological spin textures in {$1T$-phase Janus magnets: Interplay between Dzyaloshinskii-Moriya interaction,} magnetic frustration, and isotropic higher-order interactions},
  author = {Li, Peng and Yu, Dongxing and Liang, Jinghua and Ga, Yonglong and Yang, Hongxin},
  journal = {Phys. Rev. B},
  volume = {107},
  issue = {5},
  pages = {054408},
  numpages = {14},
  year = {2023},
  month = {Feb},
  publisher = {American Physical Society},
  doi = {10.1103/PhysRevB.107.054408}
}

@article{dzyaloshinskii1957,
  title={Thermodynamic theory of weak ferromagnetism in antiferromagnetic substances},
  author={Dzyaloshinskii, IE and others},
  journal={Sov. Phys. JETP},
  volume={5},
  number={6},
  pages={1259--1272},
  year={1957}
}

@article{moriya1960,
  title = {New Mechanism of Anisotropic Superexchange Interaction},
  author = {Moriya, T\^oru},
  journal = {Phys. Rev. Lett.},
  volume = {4},
  issue = {5},
  pages = {228--230},
  numpages = {0},
  year = {1960},
  month = {Mar},
  publisher = {American Physical Society},
  doi = {10.1103/PhysRevLett.4.228}
}

@Article{bode2007,
author={Bode, M.
and Heide, M.
and von Bergmann, K.
and Ferriani, P.
and Heinze, S.
and Bihlmayer, G.
and Kubetzka, A.
and Pietzsch, O.
and Bl{\"u}gel, S.
and Wiesendanger, R.},
title={Chiral magnetic order at surfaces driven by inversion asymmetry},
journal={Nature},
year={2007},
month={May},
day={01},
volume={447},
number={7141},
pages={190-193},
issn={1476-4687},
doi={10.1038/nature05802}
}

@Article{leonov2015,
author={Leonov, A. O.
and Mostovoy, M.},
title={Multiply periodic states and isolated skyrmions in an anisotropic frustrated magnet},
journal={Nature Communications},
year={2015},
month={Sep},
day={23},
volume={6},
number={1},
pages={8275},
issn={2041-1723},
doi={10.1038/ncomms9275}
}

@article{lin2016,
  title = {Ginzburg-{Landau} theory for skyrmions in inversion-symmetric magnets with competing interactions},
  author = {Lin, Shi-Zeng and Hayami, Satoru},
  journal = {Phys. Rev. B},
  volume = {93},
  issue = {6},
  pages = {064430},
  numpages = {16},
  year = {2016},
  month = {Feb},
  publisher = {American Physical Society},
  doi = {10.1103/PhysRevB.93.064430}
}

@Article{malottki2017,
author={von Malottki, S.
and Dup{\'e}, B.
and Bessarab, P. F.
and Delin, A.
and Heinze, S.},
title={Enhanced skyrmion stability due to exchange frustration},
journal={Scientific Reports},
year={2017},
month={Sep},
day={26},
volume={7},
number={1},
pages={12299},
issn={2045-2322},
doi={10.1038/s41598-017-12525-x}
}

@Article{zhang2017,
author={Zhang, Xichao
and Xia, Jing
and Zhou, Yan
and Liu, Xiaoxi
and Zhang, Han
and Ezawa, Motohiko},
title={Skyrmion dynamics in a frustrated ferromagnetic film and current-induced helicity locking-unlocking transition},
journal={Nature Communications},
year={2017},
month={Nov},
day={23},
volume={8},
number={1},
pages={1717},
issn={2041-1723},
doi={10.1038/s41467-017-01785-w}
}

@article{desplat2019,
  title = {Paths to annihilation of first- and second-order (anti)skyrmions via (anti)meron nucleation on the frustrated square lattice},
  author = {Desplat, L. and Kim, J.-V. and Stamps, R. L.},
  journal = {Phys. Rev. B},
  volume = {99},
  issue = {17},
  pages = {174409},
  numpages = {8},
  year = {2019},
  month = {May},
  publisher = {American Physical Society},
  doi = {10.1103/PhysRevB.99.174409}
}

@Article{Goerzen2023,
author={Goerzen, Moritz A.
and von Malottki, Stephan
and Meyer, Sebastian
and Bessarab, Pavel F.
and Heinze, Stefan},
title={Lifetime of coexisting sub-10{\thinspace}nm zero-field skyrmions and antiskyrmions},
journal={npj Quantum Materials},
year={2023},
month={Oct},
day={07},
volume={8},
number={1},
pages={54},
issn={2397-4648},
doi={10.1038/s41535-023-00586-3}
}

@Article{paul2020,
author={Paul, Souvik
and Haldar, Soumyajyoti
and von Malottki, Stephan
and Heinze, Stefan},
title={Role of higher-order exchange interactions for skyrmion stability},
journal={Nature Communications},
year={2020},
month={Sep},
day={21},
volume={11},
number={1},
pages={4756},
issn={2041-1723},
doi={10.1038/s41467-020-18473-x}
}

@article{kurz2001,
  title = {Three-Dimensional Spin Structure on a Two-Dimensional Lattice: {Mn/Cu(111)}},
  author = {Kurz, Ph. and Bihlmayer, G. and Hirai, K. and Bl\"ugel, S.},
  journal = {Phys. Rev. Lett.},
  volume = {86},
  issue = {6},
  pages = {1106--1109},
  numpages = {0},
  year = {2001},
  month = {Feb},
  publisher = {American Physical Society},
  doi = {10.1103/PhysRevLett.86.1106}
}

@article{koebler1996,
author={K{\"o}bler, U.
and Mueller, R.
and Smardz, L.
and Maier, D.
and Fischer, K.
and Olefs, B.
and Zinn, W.},
title={Biquadratic exchange interactions in the Europium monochalcogenides},
journal={Zeitschrift f{\"u}r Physik B Condensed Matter},
year={1996},
month={Dec},
day={01},
volume={100},
number={4},
pages={497-506},
issn={1431-584X},
doi={10.1007/s002570050153}
}

@article{koebler2001,
author = {K\"{o}bler ,Ulrich and Hoser ,A. and Englich ,J. and Snezhko ,A. and Kawakami ,M. and Beyss ,M. and Fischer ,K.},
title = {On the Failure of the Bloch–Kubo–Dyson Spin Wave Theory},
journal = {Journal of the Physical Society of Japan},
volume = {70},
number = {10},
pages = {3089-3097},
year = {2001},
doi = {10.1143/JPSJ.70.3089}
}

@article{mryasov1996,
    author = {Mryasov, O. N. and Freeman, A. J. and Liechtenstein, A. I.},
    title = {Theory of non‐Heisenberg exchange: Results for localized and itinerant magnets},
    journal = {Journal of Applied Physics},
    volume = {79},
    number = {8},
    pages = {4805-4807},
    year = {1996},
    month = {04},
    issn = {0021-8979},
    doi = {10.1063/1.361678}
}

@article{Lounis2010,
  title = {Mapping the magnetic exchange interactions from first principles: Anisotropy anomaly and application to {Fe}, {Ni}, and {Co}},
  author = {Lounis, Samir and Dederichs, Peter H.},
  journal = {Phys. Rev. B},
  volume = {82},
  issue = {18},
  pages = {180404},
  numpages = {4},
  year = {2010},
  month = {Nov},
  publisher = {American Physical Society},
  doi = {10.1103/PhysRevB.82.180404}
}

@article{yoshida2012,
  title = {Conical Spin-Spiral State in an Ultrathin Film Driven by Higher-Order Spin Interactions},
  author = {Yoshida, Y. and Schr\"oder, S. and Ferriani, P. and Serrate, D. and Kubetzka, A. and von Bergmann, K. and Heinze, S. and Wiesendanger, R.},
  journal = {Phys. Rev. Lett.},
  volume = {108},
  issue = {8},
  pages = {087205},
  numpages = {4},
  year = {2012},
  month = {Feb},
  publisher = {American Physical Society},
  doi = {10.1103/PhysRevLett.108.087205}
}

@article{Kroenlein2018,
  title = {{Magnetic Ground State Stabilized by Three-Site Interactions: $\mathrm{Fe}/\mathrm{Rh}(111)$}},
  author = {Kr\"onlein, Andreas and Schmitt, Martin and Hoffmann, Markus and Kemmer, Jeannette and Seubert, Nicolai and Vogt, Matthias and K\"uspert, Julia and B\"ohme, Markus and Alonazi, Bandar and K\"ugel, Jens and Albrithen, Hamad A. and Bode, Matthias and Bihlmayer, Gustav and Bl\"ugel, Stefan},
  journal = {Phys. Rev. Lett.},
  volume = {120},
  issue = {20},
  pages = {207202},
  numpages = {5},
  year = {2018},
  month = {May},
  publisher = {American Physical Society},
  doi = {10.1103/PhysRevLett.120.207202}
}

@article{Romming2018,
  title = {{Competition of Dzyaloshinskii-Moriya and Higher-Order Exchange Interactions in $\mathrm{Rh}/\mathrm{Fe}$ Atomic Bilayers on Ir(111)}},
  author = {Romming, Niklas and Pralow, Henning and Kubetzka, Andr\'e and Hoffmann, Markus and von Malottki, Stephan and Meyer, Sebastian and Dup\'e, Bertrand and Wiesendanger, Roland and von Bergmann, Kirsten and Heinze, Stefan},
  journal = {Phys. Rev. Lett.},
  volume = {120},
  issue = {20},
  pages = {207201},
  numpages = {6},
  year = {2018},
  month = {May},
  publisher = {American Physical Society},
  doi = {10.1103/PhysRevLett.120.207201}
}

@article{Spethmann2020,
  title = {Discovery of Magnetic Single- and Triple-$\mathbf{q}$ States in $\mathrm{Mn}/\mathrm{Re}(0001)$},
  author = {Spethmann, Jonas and Meyer, Sebastian and von Bergmann, Kirsten and Wiesendanger, Roland and Heinze, Stefan and Kubetzka, Andr\'e},
  journal = {Phys. Rev. Lett.},
  volume = {124},
  issue = {22},
  pages = {227203},
  numpages = {6},
  year = {2020},
  month = {Jun},
  publisher = {American Physical Society},
  doi = {10.1103/PhysRevLett.124.227203}
}

@article{gutzeit2021,
  title = {Trends of higher-order exchange interactions in transition metal trilayers},
  author = {Gutzeit, Mara and Haldar, Soumyajyoti and Meyer, Sebastian and Heinze, Stefan},
  journal = {Phys. Rev. B},
  volume = {104},
  issue = {2},
  pages = {024420},
  numpages = {17},
  year = {2021},
  month = {Jul},
  publisher = {American Physical Society},
  doi = {10.1103/PhysRevB.104.024420}
}

@article{takahashi1977,
doi = {10.1088/0022-3719/10/8/031},
year = {1977},
month = {apr},
publisher = {},
volume = {10},
number = {8},
pages = {1289},
author = {M Takahashi},
title = {Half-filled {Hubbard} model at low temperature},
journal = {Journal of Physics C: Solid State Physics}
}

@article{macdonald1988,
  title = {$\frac{t}{U}$ expansion for the {Hubbard} model},
  author = {MacDonald, A. H. and Girvin, S. M. and Yoshioka, D.},
  journal = {Phys. Rev. B},
  volume = {37},
  issue = {16},
  pages = {9753--9756},
  numpages = {0},
  year = {1988},
  month = {Jun},
  publisher = {American Physical Society},
  doi = {10.1103/PhysRevB.37.9753}
}

@article{hoffmann2020,
  title = {Systematic derivation of realistic spin models for beyond-{Heisenberg} solids},
  author = {Hoffmann, Markus and Bl\"ugel, Stefan},
  journal = {Phys. Rev. B},
  volume = {101},
  issue = {2},
  pages = {024418},
  numpages = {14},
  year = {2020},
  month = {Jan},
  publisher = {American Physical Society},
  doi = {10.1103/PhysRevB.101.024418}
}

@article{malottki2019,
  title = {Skyrmion lifetime in ultrathin films},
  author = {von Malottki, Stephan and Bessarab, Pavel F. and Haldar, Soumyajyoti and Delin, Anna and Heinze, Stefan},
  journal = {Phys. Rev. B},
  volume = {99},
  issue = {6},
  pages = {060409},
  numpages = {5},
  year = {2019},
  month = {Feb},
  publisher = {American Physical Society},
  doi = {10.1103/PhysRevB.99.060409}
}

@Article{Varentcova2020,
author={Varentcova, Anastasiia S.
and von Malottki, Stephan
and Potkina, Maria N.
and Kwiatkowski, Grzegorz
and Heinze, Stefan
and Bessarab, Pavel F.},
title={Toward room-temperature nanoscale skyrmions in ultrathin films},
journal={npj Computational Materials},
year={2020},
month={Dec},
day={15},
volume={6},
number={1},
pages={193},
issn={2057-3960},
doi={10.1038/s41524-020-00453-w}
}

@PhDthesis{goerzen2024thesis,
  title={Thermal equilibrium and stability of complementary topological solitons in two-dimensional magnets},
  author={Goerzen, Moritz A.},
  year={2024},
  school={University of Kiel}
}

@Article{Hagemeister2015,
author={Hagemeister, J.
and Romming, N.
and von Bergmann, K.
and Vedmedenko, E. Y.
and Wiesendanger, R.},
title={Stability of single skyrmionic bits},
journal={Nature Communications},
year={2015},
month={Oct},
day={14},
volume={6},
number={1},
pages={8455},
issn={2041-1723},
doi={10.1038/ncomms9455}
}

@article{wild2017,
author = {Johannes Wild  and Thomas N. G. Meier  and Simon Pöllath  and Matthias Kronseder  and Andreas Bauer  and Alfonso Chacon  and Marco Halder  and Marco Schowalter  and Andreas Rosenauer  and Josef Zweck  and Jan Müller  and Achim Rosch  and Christian Pfleiderer  and Christian H. Back },
title = {Entropy-limited topological protection of skyrmions},
journal = {Science Advances},
volume = {3},
number = {9},
pages = {e1701704},
year = {2017},
doi = {10.1126/sciadv.1701704},
eprint = {https://www.science.org/doi/pdf/10.1126/sciadv.1701704}}

@article{desplat2018,
  title = {Thermal stability of metastable magnetic skyrmions: Entropic narrowing and significance of internal eigenmodes},
  author = {Desplat, L. and Suess, D. and Kim, J-V. and Stamps, R. L.},
  journal = {Phys. Rev. B},
  volume = {98},
  issue = {13},
  pages = {134407},
  numpages = {13},
  year = {2018},
  month = {Oct},
  publisher = {American Physical Society},
  doi = {10.1103/PhysRevB.98.134407}
}

@article{haldar2018,
  title = {First-principles prediction of sub-10-nm skyrmions in {Pd/Fe bilayers on Rh(111)}},
  author = {Haldar, Soumyajyoti and von Malottki, Stephan and Meyer, Sebastian and Bessarab, Pavel F. and Heinze, Stefan},
  journal = {Phys. Rev. B},
  volume = {98},
  issue = {6},
  pages = {060413},
  numpages = {6},
  year = {2018},
  month = {Aug},
  publisher = {American Physical Society},
  doi = {10.1103/PhysRevB.98.060413}
}

@article{beyer2025,
  title = {Bilayer triple-Q state driven by interlayer higher-order exchange interactions},
  author = {Beyer, Bjarne and Gutzeit, Mara and Drevelow, Tim and Schwermer, Isabel and Haldar, Soumyajyoti and Heinze, Stefan},
  journal = {Phys. Rev. B},
  volume = {112},
  issue = {9},
  pages = {094430},
  numpages = {23},
  year = {2025},
  month = {Sep},
  publisher = {American Physical Society},
  doi = {10.1103/ys1g-8597}
}

@article{dupe2014,
author={Dup{\'e}, Bertrand
and Hoffmann, Markus
and Paillard, Charles
and Heinze, Stefan},
title={Tailoring magnetic skyrmions in ultra-thin transition metal films},
journal={Nature Communications},
year={2014},
month={Jun},
day={04},
volume={5},
number={1},
pages={4030},
issn={2041-1723},
doi={10.1038/ncomms5030}
}

@misc{malottki2025,
      title={Eigenmode following for direct entropy calculation and characterization of magnetic systems}, 
      author={Stephan von Malottki and Moritz A. Goerzen and Hendrik Schrautzer and Pavel F. Bessarab and Stefan Heinze},
      year={2025},
    journal={arXiv preprint arXiv:2509.10661}
}

@article{bogdanov1994,
author = {Bocdanov, A. and Hubert, A.},
title = {The Properties of Isolated Magnetic Vortices},
journal = {physica status solidi (b)},
volume = {186},
number = {2},
pages = {527-543},
doi = {https://doi.org/10.1002/pssb.2221860223},
year = {1994}
}

@article{muller2018,
  title = {Duplication, Collapse, and Escape of Magnetic Skyrmions Revealed Using a Systematic Saddle Point Search Method},
  author = {M\"uller, Gideon P. and Bessarab, Pavel F. and Vlasov, Sergei M. and Lux, Fabian and Kiselev, Nikolai S. and Bl\"ugel, Stefan and Uzdin, Valery M. and J\'onsson, Hannes},
  journal = {Phys. Rev. Lett.},
  volume = {121},
  issue = {19},
  pages = {197202},
  numpages = {6},
  year = {2018},
  month = {Nov},
  publisher = {American Physical Society},
  doi = {10.1103/PhysRevLett.121.197202}
}

@Article{bessarab2018,
author={Bessarab, Pavel F.
and M{\"u}ller, Gideon P.
and Lobanov, Igor S.
and Rybakov, Filipp N.
and Kiselev, Nikolai S.
and J{\'o}nsson, Hannes
and Uzdin, Valery M.
and Bl{\"u}gel, Stefan
and Bergqvist, Lars
and Delin, Anna},
title={Lifetime of racetrack skyrmions},
journal={Scientific Reports},
year={2018},
month={Feb},
day={21},
volume={8},
number={1},
pages={3433},
issn={2045-2322},
doi={10.1038/s41598-018-21623-3}
}

@article{bessarab2012,
  title = {Harmonic transition-state theory of thermal spin transitions},
  author = {Bessarab, Pavel F. and Uzdin, Valery M. and J\'onsson, Hannes},
  journal = {Phys. Rev. B},
  volume = {85},
  issue = {18},
  pages = {184409},
  numpages = {4},
  year = {2012},
  month = {May},
  publisher = {American Physical Society},
  doi = {10.1103/PhysRevB.85.184409}
}

@article{schrautzer2022,
  title = {Effects of interlayer exchange on collapse mechanisms and stability of magnetic skyrmions},
  author = {Schrautzer, Hendrik and von Malottki, Stephan and Bessarab, Pavel F. and Heinze, Stefan},
  journal = {Phys. Rev. B},
  volume = {105},
  issue = {1},
  pages = {014414},
  numpages = {21},
  year = {2022},
  month = {Jan},
  publisher = {American Physical Society},
  doi = {10.1103/PhysRevB.105.014414}
}

@PhDThesis{Kurz2000,
  title = {{Non-Collinear Magnetism at Surfaces and in Ultrathin Films}},
  author = {Kurz, P.},
  school = {Rheinisch-Westfälische Technische Hochschule Aachen},
  year = {2000}
  }

@article{bessarab2015,
title = {Method for finding mechanism and activation energy of magnetic transitions, applied to skyrmion and antivortex annihilation},
journal = {Computer Physics Communications},
volume = {196},
pages = {335-347},
year = {2015},
issn = {0010-4655},
doi = {https://doi.org/10.1016/j.cpc.2015.07.001},
author = {Pavel F. Bessarab and Valery M. Uzdin and Hannes Jónsson}
}

@article{schrautzer2025,
  title = {Identification of mechanisms of magnetic transitions using an efficient method for converging on first-order saddle points},
  author = {Schrautzer, Hendrik and Sallermann, Moritz and Bessarab, Pavel F. and J\'onsson, Hannes},
  journal = {Phys. Rev. B},
  volume = {112},
  issue = {10},
  pages = {104433},
  numpages = {20},
  year = {2025},
  month = {Sep},
  publisher = {American Physical Society},
  doi = {10.1103/z673-hhnp}
}

@article{arya2025new,
  title={A new skyrmion topological transition driven by higher-order exchange interactions in {Janus MnSeTe}},
  author={Arya, Megha and Goerzen, Moritz A and Calmels, Lionel and Arras, R{\'e}mi and Haldar, Soumyajyoti and Heinze, Stefan and Li, Dongzhe},
  journal={arXiv preprint arXiv:2509.10661},
  year={2025}
}

@article{gutzeit2022,
author={Gutzeit, Mara
and Kubetzka, Andr{\'e}
and Haldar, Soumyajyoti
and Pralow, Henning
and Goerzen, Moritz A.
and Wiesendanger, Roland
and Heinze, Stefan
and von Bergmann, Kirsten},
title={Nano-scale collinear multi-Q states driven by higher-order interactions},
journal={Nature Communications},
year={2022},
month={Sep},
day={30},
volume={13},
number={1},
pages={5764},
issn={2041-1723},
doi={10.1038/s41467-022-33383-w}
}

\end{document}